\documentclass[9pt,twocolumn,twoside]{pnas-new}

\newcommand{\pder}[2]{\frac{\partial#1}{\partial#2}} 
\newcommand{\pderi}[2]{\partial#1 / \partial#2} 

\templatetype{pnasresearcharticle} 

\title{Velocity transformation for compressible wall-bounded turbulent flows with and without heat transfer}

\author[a]{Kevin Patrick Griffin}
\author[a,$\dagger$,*]{Lin Fu} 
\author[a,*]{Parviz Moin}
\affil[a]{Center for Turbulence Research, Stanford University, Stanford, CA 94305}

\leadauthor{Griffin} 

\significancestatement{Near a solid wall, the mean velocity of an incompressible fluid is a nearly universal function of the distance from the wall when appropriately transformed (non-dimensionalized). For high-speed compressible flows, due to the wall-normal variations of density and viscosity, it has not been established how to transform a dimensional velocity profile into a universal profile. We propose a transformation that is more accurate and more broadly applicable than existing approaches. This improvement results from deploying different physical arguments for the two critical sub-domains of a boundary layer. The transformation can be used to extend incompressible turbulence models to compressible turbulent flows, such as those encountered by vehicles for high-speed transportation and planetary reentry.}

\authorcontributions{K.P.G. and L.F. initialized and designed the research. P.M. advised the research. K.P.G. mainly performed the research. L.F. and P.M. contributed new ideas and turbulence perspectives. K.P.G. and L.F. analyzed the data and wrote the paper. P.M. reviewed and improved the paper.}
\authordeclaration{The authors declare no conflicts of interest.}
\equalauthors{\textsuperscript{$\dagger$}Current affiliations: Department of Mechanical and Aerospace Engineering, The Hong Kong University of Science and Technology, Clear Water Bay, Kowloon, Hong Kong; Department of Mathematics, The Hong Kong University of Science and Technology, Clear Water Bay, Kowloon, Hong Kong}
\correspondingauthor{\textsuperscript{*}To whom correspondence should be addressed. E-mail: linfu@ust.hk (LF), moin@stanford.edu (PM)}

\keywords{turbulence $|$ wall modeling $|$ law of the wall $|$ mean velocity profile $|$ Mach invariance} 

\begin{abstract}
In this work, a transformation, which maps the mean velocity profiles of compressible wall-bounded turbulent flows to the incompressible law of the wall is proposed. Unlike existing approaches, the proposed transformation successfully collapses, without specific tuning, numerical simulation data from fully developed channel and pipe flows, and boundary layers with or without heat transfer. In all these cases, the transformation is successful across the entire inner layer of the boundary layer (including the viscous sublayer, buffer layer, and logarithmic layer), recovers the asymptotically exact near-wall behavior in the viscous sublayer, and is consistent with the near balance of turbulence production and dissipation in the logarithmic region of the boundary layer. The performance of the transformation is verified for compressible wall-bounded flows with edge Mach numbers ranging from 0 to 15 and friction Reynolds numbers ranging from 200 to 2000. Based on physical arguments, we show that such a general transformation exists for compressible wall-bounded turbulence regardless of the wall thermal condition.
\end{abstract}

\dates{This manuscript was compiled on \today}
\doi{\url{www.pnas.org/cgi/doi/10.1073/pnas.XXXXXXXXXX}}

\begin{document}

\maketitle
\thispagestyle{firststyle}
\ifthenelse{\boolean{shortarticle}}{\ifthenelse{\boolean{singlecolumn}}{\abscontentformatted}{\abscontent}}{}

\dropcap{I}t is well known that incompressible wall-bounded turbulent flows at high Reynolds numbers exhibit nearly-universal mean streamwise velocity profiles versus the wall-normal coordinate. This collapse is achieved through a non-dimensionalization with respect to the friction velocity and the kinematic viscosity. However, compressible flows with non-negligible freestream Mach numbers do not directly obey this law of the wall. Morkovin \cite{Morkovin1962} hypothesizes that, as long as the turbulent Mach number (based on the root-mean-square fluctuating velocity) is sufficiently small, compressible wall-bounded flows can be mapped onto the incompressible counterparts by accounting for the variation in mean properties (e.g., density, viscosity, etc.). Such a transformation would not only be of theoretical interest but also of great practical importance for reduced-order turbulence modeling, since it would allow established incompressible wall models to be readily applied to compressible flows. Since the pioneering work of van Driest \cite{VanDriest1951}, many variants have been proposed over the past decades, e.g., \cite{Zhang2012,Trettel2016,Patel2016,Volpiani2020}, but none of the existing transformations is generally applicable even to canonical compressible wall-bounded flows. In this work, the strengths and limitations of existing transformations are first analyzed. Then, a generally applicable transformation is proposed, validated, and compared with existing approaches for a wide range of compressible wall-bounded turbulent flows.


The first successful velocity transformation is developed by van Driest \cite{VanDriest1951} by asserting that compressible wall bounded turbulence obeys Prandtl's incompressible mixing length assumption. This implies that the non-dimensional mean shear,
$S_{VD}^+[y^+] = \pderi{U_{VD}^+}{y^+} = \sqrt{\rho^+} \pderi{U^+}{y^+}$,
is a Mach-number-independent function. Throughout this paper, the superscript $^+$ indicates a non-dimensionalization by the friction velocity $u_\tau = \sqrt{\tau_w/\rho_w}$, the viscous length scale $\delta_v=\mu_w/(u_\tau \rho_w)$, and $\rho_w$, where $\tau_w$, $\rho_w$, and $\mu_w$ are the shear stress, density, and dynamic viscosity evaluated on the wall, respectively. For instance, $U^+=U/u_\tau$, $y^+ = y/\delta_v$, $\mu^+ = \mu/\mu_w$, $\rho^+=\rho/\rho_w$, etc. 

If a particular non-dimensionalization of the mean shear profile is indeed a Mach-number-invariant function, then it can be integrated to recover the law of the wall, and this formula can be referred to as a successful compressible velocity transformation.
For the van Driest transformation \cite{VanDriest1951}, Mach invariance is achieved in adiabatic boundary layer flows, since the transformed velocity profiles collapse well to incompressible reference data (see Fig.~3(a) in \cite{Pirozzoli2011}). However, in diabatic (with heat transfer) boundary layers and channel flows, the transformation fails and the incompressible law of the wall is not recovered \cite{Zhang2018,Volpiani2018,Volpiani2020a,Fu2021,Modesti2016,Trettel2016,Yao2020}. 


Zhang et al. \cite{Zhang2012} propose replacing van Driest's mixing length assumption with the more general proposition of turbulence equilibrium. Specifically, $P/\epsilon \approx 1$, where $P$ and $\epsilon$ denote the turbulence production and dissipation, respectively. 
This is a statement that the turbulence cascade is in equilibrium across the entire inner layer of the boundary layer. Furthermore, the improved model assumes that the turbulent shear stress $\tau_R^{+} =  -\overline{\rho}\widetilde{u''v''}/\tau_w$ \cite{Morkovin1962} and $\epsilon^{+}/\mu^{+}$ are Mach-independent functions of the non-dimensional wall-normal coordinate $y^+$, where $\widetilde{\cdot}$ denotes the density weighted average (Favre average) and $\cdot''$ denotes the fluctuation about this average. Recalling that $P = -\overline{\rho} \widetilde{u''v''} \pderi{U}{y}$ (where $U$ is the Favre averaged velocity henceforth), these assumptions imply that $(1/\mu^+)  (\pderi{U^+}{y^+}) = S_1^+[y^+]$, where $S_1^+$ is some Mach-independent function. Note that, throughout this work, brackets $[\cdot]$ denote functional dependence.

However, Zhang et al. \cite{Zhang2012} further report that $S_1^+$ is not Mach invariant in the viscous sublayer, so it is inaccurate to directly integrate $S_1^+$ as a velocity transformation. Similar to Huang and Coleman \cite{Huang1994}, they proceed by deriving the mixing length that is implied by $S_1^+$, assuming the mixing length is Mach invariant across the entire inner layer of the boundary layer, and further assuming the existence of a constant stress layer near the wall. These assumptions imply the following transformation
$S_Z^+[y^+] = \left( -S_1^{+,2} + \sqrt{ S_1^{+,2}/4 + (1-\mu^{+2}S_1^+)} \right) S_1^+/(1 - \mu^{+2}S_1^+)$ \cite{Zhang2012}.
Zhang et al. recommend the use of $S_Z^+$ instead of $S_1^+$. Although the $S_Z^+$ transformation performs well in adiabatic boundary layers (see Fig.~4 in \cite{Zhang2012}), the performance in channel flows, pipe flows, and diabatic boundary layers is poor, which will be documented later in this paper.

The observation that $S_1^+$ varies with the Mach number in the viscous sublayer should be expected since the derivation of the expression for $S_1^+$ is based on the equilibrium assumption of $P/\epsilon \approx 1$, which is not justifiable in the viscous sublayer and buffer layer. In this work, separate physical arguments will be invoked in those regions.

\section*{Transformation based on turbulence quasi-equilibrium}

Before considering the viscous sublayer, we first generalize the log layer arguments of Zhang et al. \cite{Zhang2012} to diabatic flows.
Huang et al. \cite{Huang1995} show that, for cooled walls, the semi-local velocity scale $u_{sl}=\sqrt{\tau_w/\rho[y]}$ and length scale $\ell_{sl} = \nu[y]/u_{sl}$ are more appropriate for mapping the wall-normal distributions of turbulence statistics to their incompressible counterparts. In addition, Coleman et al. \cite{Coleman1995} numerically show that a good collapse is obtained when plotting the turbulence statistics versus the semi-local wall-normal coordinate $y^* = y / \ell_{sl}$ for different diabatic supersonic channel flows. 

\begin{figure*}
	\centering
	\includegraphics[width=0.45\linewidth]{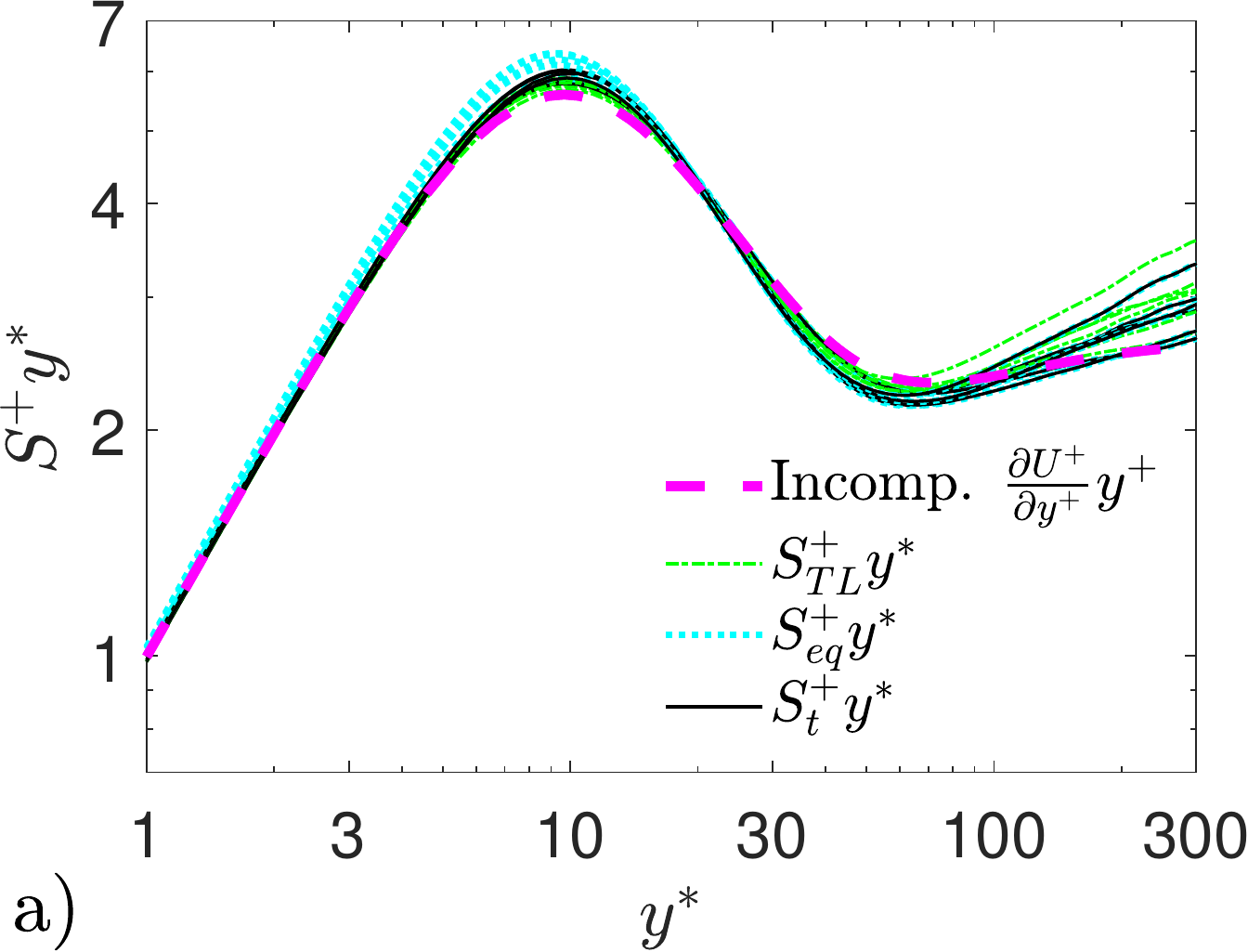}
	\includegraphics[width=0.45\linewidth]{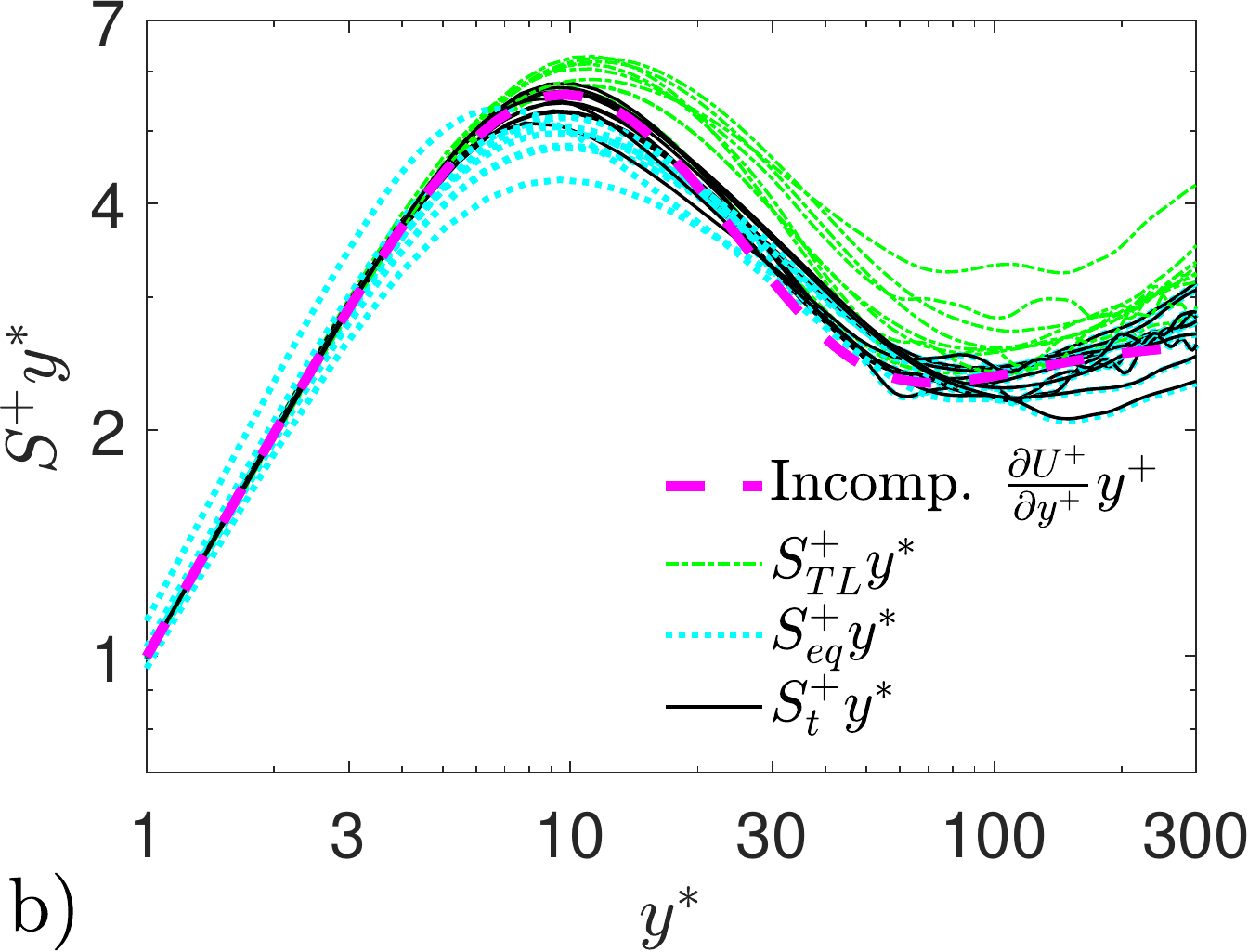}
	\caption{Three non-dimensionalizations of the mean shear (that of Trettel-Larsson in green characterized by Eq.~(\ref{eq:S_TL}), the one based on turbulence equilibrium in cyan characterized by Eq.~(\ref{eq:S_eq}), and the total-stress-based transformation in black characterized by Eq.~(\ref{eq:St_simp}))  multiplied by the semi-local wall-normal coordinate are plotted against this same coordinate.
	The pre-multiplied non-dimensional mean shear $y^+\pderi{U^+}{y^+}$ from an incompressible channel flow \cite{Lee2015} of $Re_\tau\approx 5200$ is shown for reference. 
	Profiles of channel flows \citep{Modesti2016,Trettel2016,Yao2020} with $Re_\tau^*>600$ are shown in (a) and profiles of boundary layers \citep{Pirozzoli2011,Zhang2018,Volpiani2018,Volpiani2020a,Fu2021} with $Re_\tau^* > 2000$ are shown in (b). Wake effects are likely not present since only the near-wall data within $y/\delta<0.15$ is shown, where $\delta$ is the boundary layer thickness defined in \cite{Griffin2021}.
	}
\label{fig:dU}
\end{figure*}

Following Huang et al. \cite{Huang1995} and Coleman et al. \cite{Coleman1995}, to account for the diabatic effects, the proposition by Zhang et al. \cite{Zhang2012} is generalized to be that $\tau_R^{+} = -\overline{\rho}\widetilde{u''v''}/\tau_w$ and $\epsilon^{*} = \epsilon^+/\mu^+$ are Mach-independent functions of $y^*$ instead of $y^+$, where the $^*$ superscripts denote non-dimensionalization with respect to semi-local quantities, and the non-dimensional production $P^\ddagger = \tau_R^{+} \pderi{U^+}{y^*}$ is assumed to approximately balance with the non-dimensional dissipation $\epsilon^+$. Rather than requiring a precise balance $P^\ddagger/\epsilon^{+} \approx 1$ as in Zhang et al. \cite{Zhang2012}, a strictly weaker assumption of an approximate balance (or quasi-equilibrium) is invoked, namely $P^\ddagger/\epsilon^{+}$ is a Mach-independent function of $y^*$. These assumptions imply that 
\begin{equation*}
\left(\frac{P^\ddagger}{\epsilon^+}\right)[y^*] = \frac{\tau_R^{+}[y^*] }{\epsilon^*[y^*] } \frac{\pderi{U^+}{y^*}}{\mu^+}.
\end{equation*}
The Mach invariance of $P^\ddagger/\epsilon^+$, $\tau_R^{+}$, and $\epsilon^*$ with respect to $y^*$ implies that
the following non-dimensionalization of the mean shear
\begin{equation} \label{eq:S_eq}
    S^{+}_{\mathrm{eq}}[y^*] = \frac{1}{\mu^+} \pder{U^+}{y^*} = \pder{U_{eq}^+}{y^*},
\end{equation}
is a Mach-independent function in the log region. This definition of the non-dimensional mean shear also defines the transformed velocity $U_{eq}$.
These assertions of Mach independence are validated collectively in Fig.~\ref{fig:dU}(a) and (b). These figures show distributions of the premultiplied non-dimensional mean shear $S^+ y^*$ versus the semi-local wall-normal coordinate $y^*$ for various channel and pipe flows in Fig.~\ref{fig:dU}(a) and boundary layers in Fig.~\ref{fig:dU}(b). $S^{+}_{\mathrm{eq}}$ is indeed Mach invariant since a wide range of compressible velocity profiles collapse to the incompressible reference data in the log region. $S^{+}_{\mathrm{eq}}$ is referred to as the non-dimensionalization of the mean shear based on turbulence quasi-equilibrium, and it is suitable for both adiabatic and diabatic wall-bounded flows in the log layer. Unlike the approach of Zhang et al. \cite{Zhang2012}, the mixing-length and constant-stress-layer assumptions have not been invoked. Also, this transformation will only be deployed for the Reynolds shear stresses. The viscous stresses will be handled in the following development. 

\section*{Transformation based on the viscous stress}

Next, we consider the physical arguments applicable to the viscous sublayer. Carvin et al. \cite{Carvin1988} employ the conventional assumptions for this region (i.e. they neglect the Reynolds shear stress and assume that the total shear stress is invariant with $y^+$, which together imply $\mu^+ \pderi{U^+}{y^+} \approx 1$ in the viscous sublayer) to define the compressible velocity transformation $S_{vs}^+[y^+] = \mu^+ \pderi{U^+}{y^+} = \pderi{U_{vs}^+}{y^+}$, which is only guaranteed to be Mach invariant in the viscous sublayer. Trettel-Larsson \cite{Trettel2016} and Patel et al. \cite{Patel2016} reinterpret $\mu^+ \pderi{U^+}{y^+}$ as the semi-local non-dimensionalization \cite{Huang1995,Coleman1995} of the mean shear and propose that this non-dimensionalization should be applied to the wall-normal coordinate as well, leading to the velocity transformation
\begin{equation} \label{eq:S_TL}
S_{TL}^{+}[y^*] = \mu^+ \pder{U^+}{y^+} = \pder{U_{TL}^+}{y^*},
\end{equation}
where the transformed velocity $U_{TL}^+$ is defined accordingly.

Patel et al. \cite{Patel2016} observe that, while $S_{vs}^+[y^+]$ only exhibits Mach invariance in the viscous sublayer, $S_{TL}^{+}[y^*]$ exhibits Mach invariance across the entire domain in channel flows for various prescribed viscosity profiles. Trettel and Larsson \cite{Trettel2016} provide an argument for the applicability of this transformation outside the viscous sublayer by appealing to the Mach-independence of the Reynolds shear stress \cite{Morkovin1962} and applying the constant-stress-layer assumption outside its domain of applicability, to the entire half channel. Note that the transformation has been documented to be successful across the entire half channel \cite{Trettel2016,Patel2016} even though these assumptions are not exactly valid. These assumptions imply the Mach invariance of the viscous stress $\tau_v^+ = \tau^+ - \tau_R^+$. Although the individual models for $\tau^+$ and $\tau_R^+$ are reasonably accurate in the inner layer (the errors are typically $\lesssim 10\%$ for $y/\delta < 0.1$), the resulting model for the difference of these terms can have unbounded percent error, since the terms approximately balance for $y^*>30$. This suggests that the Mach invariance of this transformation is not guaranteed when the viscous stress is small (such as in the log region).

\begin{figure*}
\centering
  \includegraphics[width=0.3\linewidth]{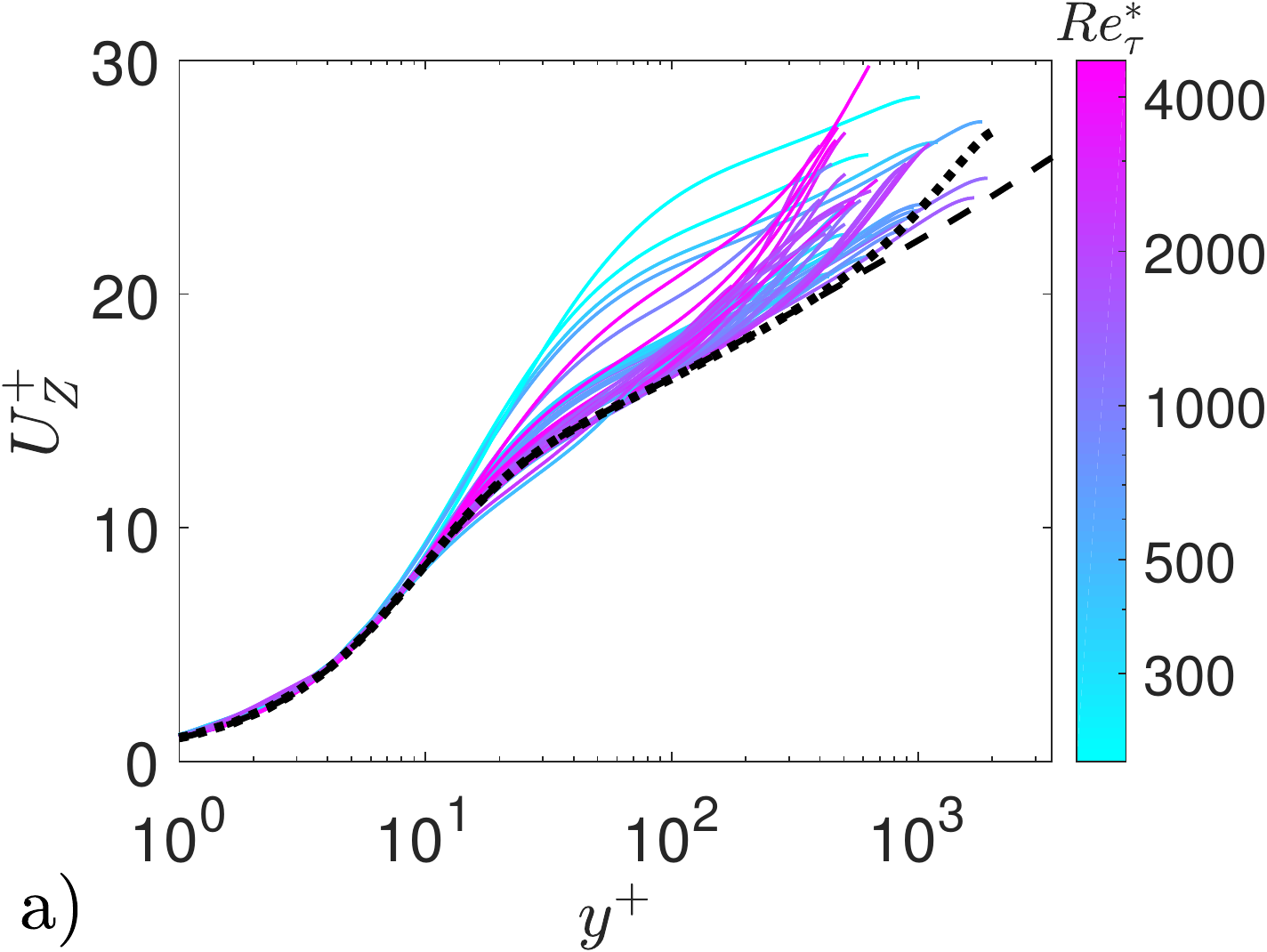}
  \includegraphics[width=0.3\linewidth]{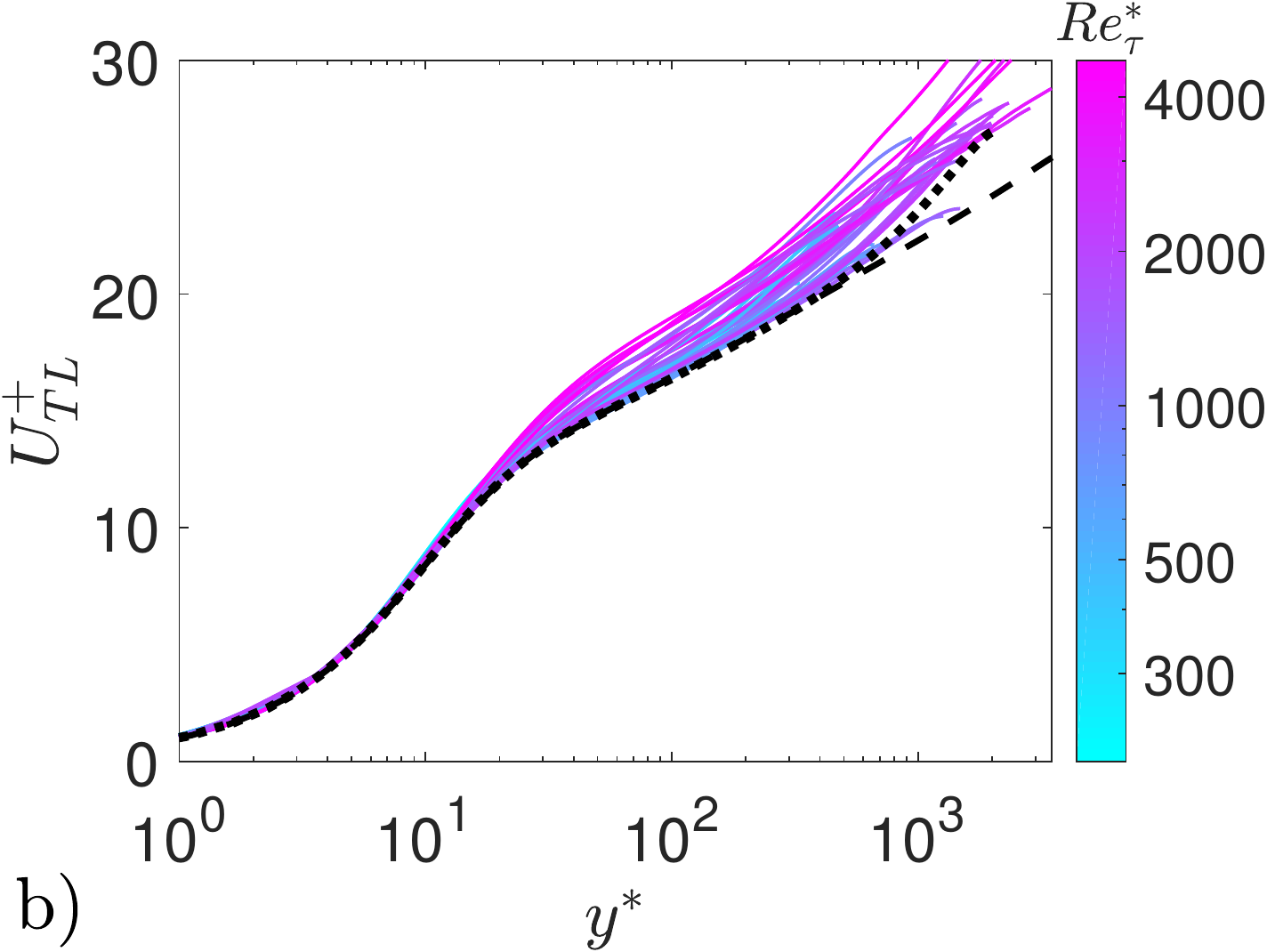}
  \linebreak
  \includegraphics[width=0.3\linewidth]{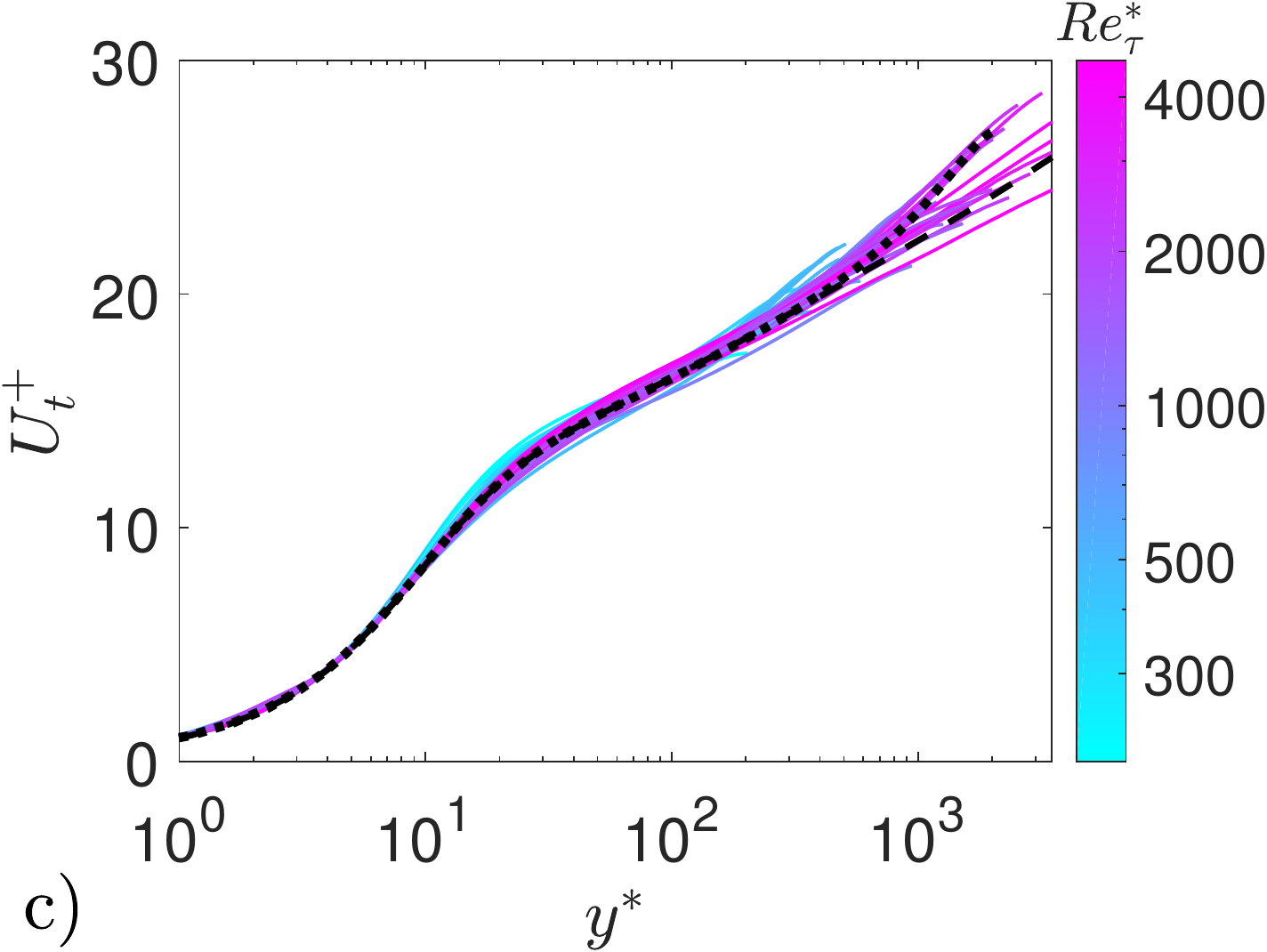}
  \includegraphics[width=0.3\linewidth]{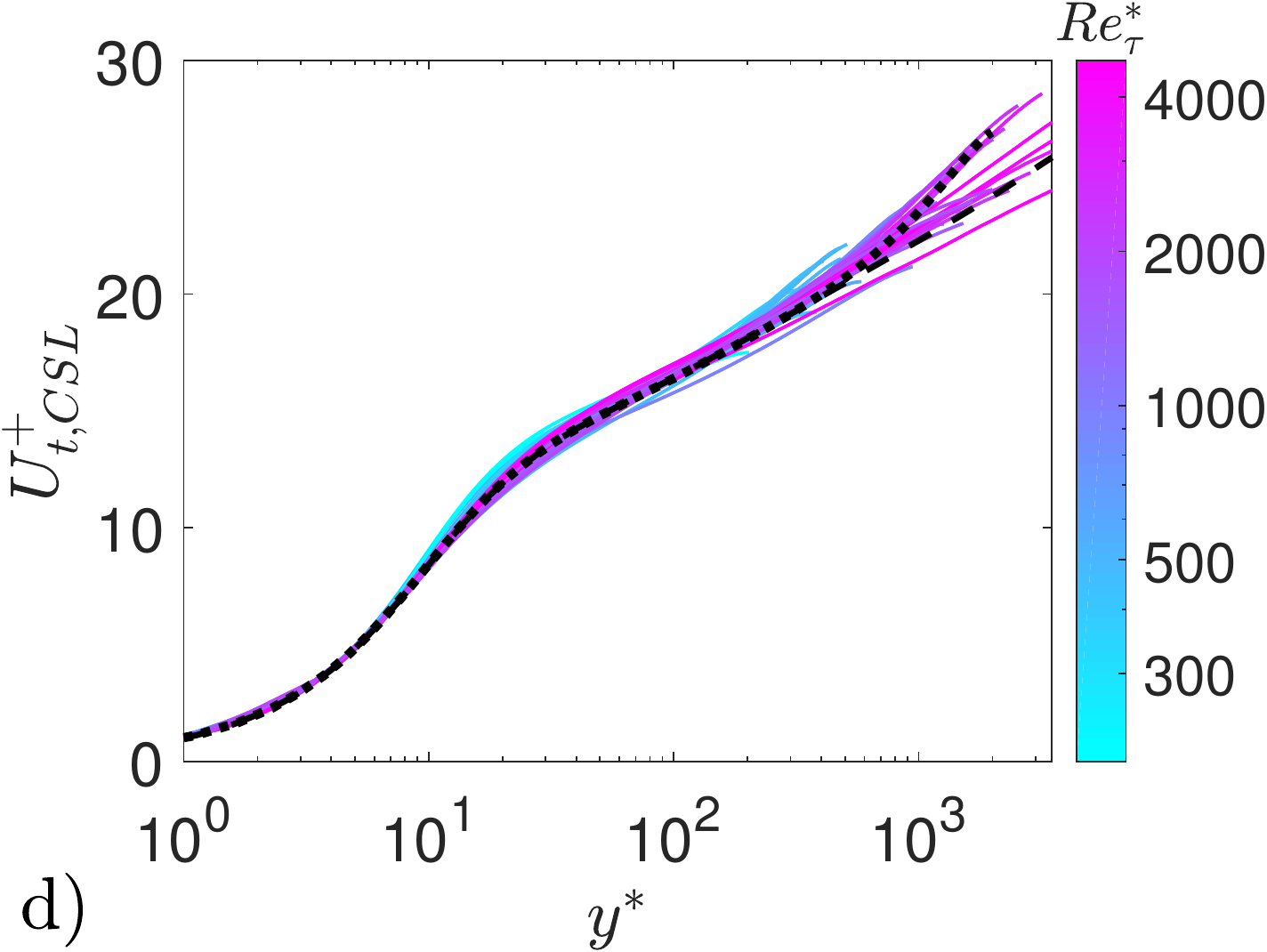}
\caption{The Zhang et al. \cite{Zhang2012} (a), Trettel-Larsson \cite{Trettel2016} (b), and present total-stress-based (c) compressible velocity transformations versus the semi-local wall-normal coordinate. An approximate version of the total-stress-based transformation based on Eq.~(\ref{eq:St_CSL}) is shown in (d) in which the constant-stress-layer assumption, i.e., $\tau^+=1$, is invoked instead of using the exact total shear stress data as in Eq.~(\ref{eq:St_simp}). The database includes: the adiabatic boundary layers \citep{Pirozzoli2011,Zhang2018,Volpiani2018}, diabatic boundary layers (with both heated and cooled walls) \citep{Zhang2018,Volpiani2018,Volpiani2020a}, channel flows \citep{Modesti2016,Trettel2016,Yao2020}, pipe flows \citep{Modesti2019}, and a diabatic turbulent boundary layer downstream of the impingement of a shockwave on a laminar boundary layer \citep{Fu2021}. The line color indicates the semi-local Reynolds number $Re_\tau^* = \delta \sqrt{\tau_w \rho}/\mu$. All data have $Re_\tau^*$ larger than 200. Incompressible channel data of Lee and Moser \cite{Lee2015} with $Re_\tau \approx 5200$ (black dashed lines) and the zero-pressure-gradient boundary layer data of Sillero et al. \cite{Sillero2013} with $Re_\tau \approx 2000$ (black dotted lines) are shown for reference.}
\label{fig:all_cases}
\end{figure*}

Prior numerical evidence supports this criticism. Although the transformation has been successful in channel flows even in the log region (see \cite{Trettel2016,Patel2016,Modesti2016,Modesti2019,Yao2020}), it has displayed some variability in the log intercepts of transformed velocity profiles from adiabatic boundary layers, and the variability is even more pronounced in diabatic boundary layers (see \cite{Zhang2018,Volpiani2018,Volpiani2020a,Fu2021}). In all of these cases, the transformation performs well in the viscous sublayer where the viscous stress $\tau_v > \tau_R$, but in boundary layers it is less successful in the log region where $\tau_R>\tau_v$. This is clearly demonstrated in Fig \ref{fig:dU}(b), where $S_{TL}^{+} y^*$ collapses to the incompressible reference in the near-wall region, but for $y^*>30$, $S_{TL}^{+}$ is overpredicted for many cases.
(Note that errors in the mean shear are cumulative when computing the transformed velocity, so even the slight, but sustained, overprediction of mean shear profiles leads to an overprediction of the log-intercept in transformed velocity profiles as shown in Fig.~\ref{fig:all_cases}(b).)
This motivates the present transformation, which treats the viscous stress and Reynolds shear stress separately using established principles of boundary layer theory.

%
\section*{Total-stress-based transformation}
Leveraging the prior analyses of the viscous sublayer and the log layer, a velocity transformation that is valid across the entire inner layer is derived by first recalling that the total shear stress is the sum of the viscous and Reynolds shear stresses, i.e., $\tau^+ = \tau_v^+ + \tau_R^+$. The total stress can be written in terms of a generalized non-dimensional mean shear $S_t^{+}$, which is defined by
\begin{equation} \label{eq:St_defn}
	\tau^+ = S_t^{+} (\frac{\tau_v^+}{S_{TL}^{+}}+\frac{\tau_R^+}{S_{eq}^{+}}).
\end{equation}
By construction, in the near-wall limit where $\tau_v^+\rightarrow\tau^+$, the relation $S_t^{+} \rightarrow S_{TL}^{+}$ holds, thus recovering the Trettel Larsson \cite{Trettel2016} transformation, which has the asymptotically correct near-wall behavior derived by Carvin et al. \cite{Carvin1988}. Meanwhile, in the log region where $\tau_R^+ \rightarrow \tau^+$, $S_t^{+} \rightarrow S_{eq}^{+}$, thus recovering the non-dimensionalization of the mean shear based on the approximate balance of turbulence production and dissipation.
Observing that Eq.~(\ref{eq:S_TL}) implies $\tau_v^+=S_{TL}^{+}$, and with $\tau_R^+ = \tau^+ - \tau_v^+$, Eq.~(\ref{eq:St_defn}) can be rearranged for $S_t^{+}$ as,
\begin{equation} \label{eq:St_simp}
    S_{t}^{+} = \frac{\tau^+ S_{eq}^{+} }{\tau^+ + S_{eq}^{+} - S_{TL}^{+}}.
\end{equation}
This equation (the presently proposed transformation) is referred to as the total-stress-based transformation since it is an overall non-dimensionalization of the mean shear that treats the viscous and Reynolds shear stresses independently using results from boundary layer theory, only in their domains of applicability.
Integration with respect to the semi-local wall-normal coordinate leads to the transformed velocity $U_t^+[y^*] = \int S_t^{+} dy^*$.

As a practical note, if the total stress profile is not available, the present transformation in Eq.~(\ref{eq:St_simp}) can be combined with a constant-stress-layer assumption, $\tau^+\approx 1$, which is suitable for the canonical flows considered in this work.
\begin{equation} \label{eq:St_CSL}
    S_{t}^+ \approx \frac{S_{eq}^+ }{1 + S_{eq}^+ - S_{TL}^+}.
\end{equation}
Comparing Fig.~\ref{fig:all_cases}(c) and Fig.~\ref{fig:all_cases}(d), for this data, there is no discernible difference between using the exact total shear stress profile and the constant-stress-layer model.

For incompressible flows, there is an established definition of the eddy viscosity $\mu_t^+[y^+] = \tau_R^+/(\pderi{U^+}{y^+})$. For compressible flows, it is an open question how to best define the compressible eddy viscosity to facilitate modeling. 
The present velocity transformation in Eq.~(\ref{eq:St_simp}) suggests the definition of the compressible eddy viscosity as follows
\begin{equation}
\mu_{t}^\ddagger = \frac{\tau_R^+}{S_{eq}^{+}} = \frac{\mu^+ \tau_R^+}{ \pderi{U^+}{y^*}}.
\end{equation}
Then, Eq.~(\ref{eq:St_defn}) implies that the total shear stress can be expressed in terms of the compressible eddy viscosity as $\tau^+ = S_t^{+} (1+\mu_t^\ddagger)$. Note that this equation also holds for the incompressible limit, $\mu_t^\ddagger \rightarrow \mu_t^+$. 

\begin{figure}
	\centering
	\includegraphics[width=0.9\linewidth]{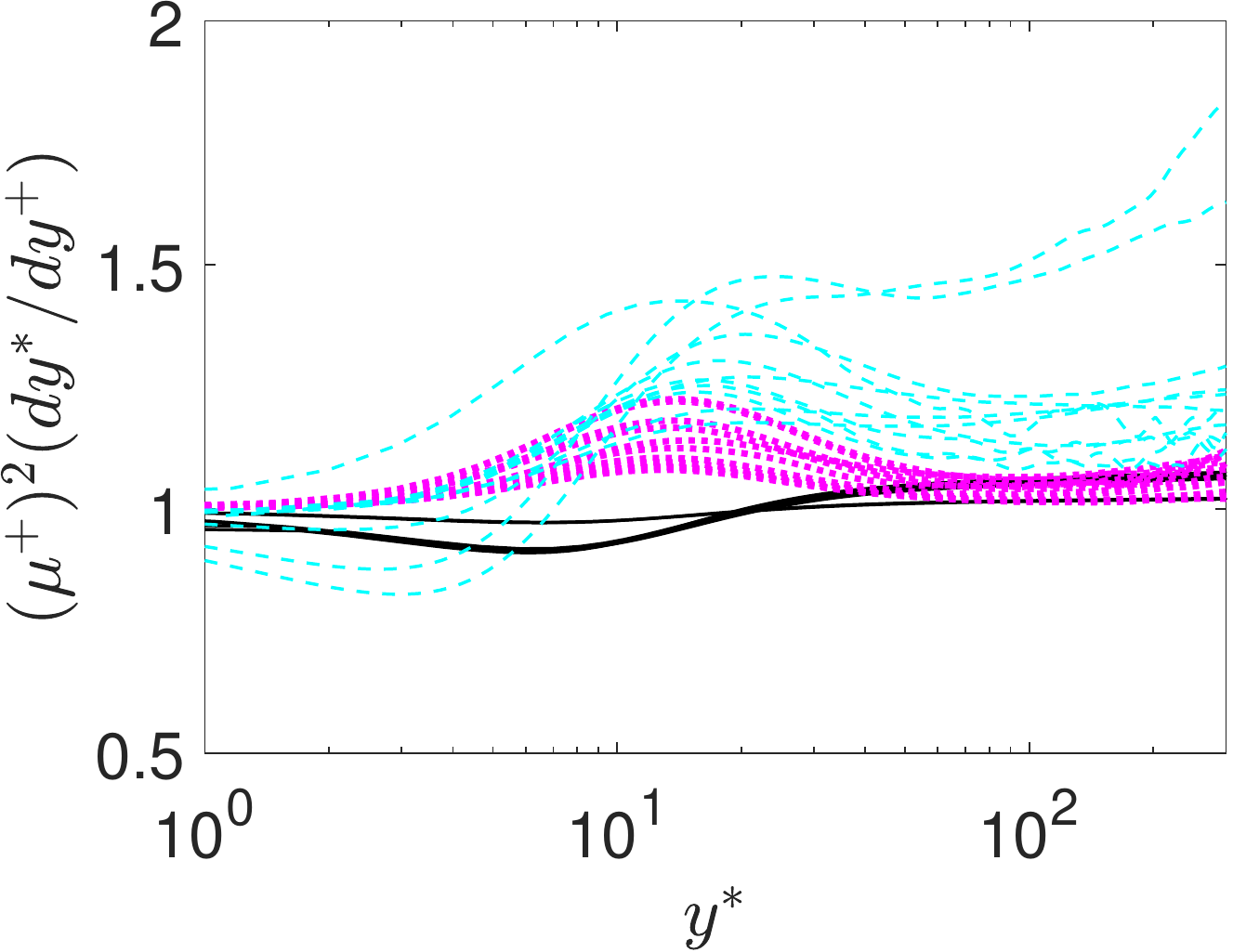}
	\caption{$(\mu^+)^{2}\pderi{y^*}{y^+}$ versus the semi-local wall-normal coordinate. Unity indicates the equivalence of the present and Trettel-Larsson \cite{Trettel2016} transformations. Profiles of channel and pipe flows (black) \citep{Modesti2016,Trettel2016,Yao2020,Modesti2019}, adiabatic boundary layers (magenta) \citep{Pirozzoli2011,Zhang2018,Volpiani2018}, and diabatic boundary layers (cyan) \citep{Zhang2018,Volpiani2018,Volpiani2020a,Fu2021} . For all cases plotted, $Re_\tau^* > 600$.
	}
\label{fig:mu2}
\end{figure}

On the other hand, a different definition of the compressible eddy viscosity is implied by the Trettel-Larsson transformation \cite{Trettel2016}. Although their formulation does not explicitly define a compressible eddy viscosity, the total stress formalism proposed above can be applied to their transformation by redefining the compressible eddy viscosity as 
\begin{equation}
\mu_{t,TL}^\ddagger = \frac{\tau_R^+}{S_{TL}^{+}} = \frac{\tau_R^+}{\mu^+ \pderi{U^+}{y^+}}.
\end{equation}
Then, the analogy to Eq.~(\ref{eq:St_defn}) is $\tau^+ = S_{TL}^{+} (1+\mu_{t,TL}^\ddagger)$, where the viscous and Reynolds shear stresses have both been written in terms of $S_{TL}^{+}$.
The fact that $\mu_t^\ddagger$ and $\mu_{t,TL}^\ddagger$ are not equivalent, in general, suggests that the semi-local accounting of mean property effects in the Reynolds shear stress is not equivalent to accounting for the change in the mean shear required to maintain turbulence in quasi-equilibrium. However, these two assumptions are equivalent when and only when the condition,
$(\mu^+)^{2} (\pderi{y^*}{y^+})=1$, is satisfied.
Fig.~\ref{fig:dU}(a) shows that for moderate $Re_\tau^*$ channel and pipe flows, the Trettel-Larsson and turbulence quasi-equilibrium transformations are equivalent. Consistently, Fig.~\ref{fig:mu2} indicates that $(\mu^+)^{2} (\pderi{y^*}{y^+}) \approx 1$ for these flows. Meanwhile, Fig.~\ref{fig:dU}(b) shows an overprediction of the mean shear by the Trettel-Larsson transformation in several boundary layer flows, which is also consistent with the departure from unity of $(\mu^+)^{2} (\pderi{y^*}{y^+})$ in Fig.~\ref{fig:mu2} for the boundary layers.
The transformed velocity profiles for the boundary layers, channel and pipe flows are shown in Fig.~\ref{fig:all_cases}. The results for the transformation of Zhang et al. \cite{Zhang2012} in Fig.~\ref{fig:all_cases}(a) do not collapse to the incompressible reference since this transformation relies on a mixing length definition and a mean shear non-dimensionalization that are only applicable to adiabatic flows. The results for the Trettel-Larsson transformation \cite{Trettel2016} in Fig.~\ref{fig:all_cases}(b) exhibit considerable scatter in the logarithmic region, which is consistent with the overprediction of the mean shear in the beginning of the log region observed in Fig.~\ref{fig:dU}(b). The total-stress-based transformation in Fig.~\ref{fig:all_cases}(c) is accurate for all flows and lies within the reference data, as the remaining scatter is mostly confined to the boundary layer wake $y/\delta>0.2$, where even the incompressible data is not universal.

In addition, two empirical velocity transformations have been developed. 1) Iyer and Malik \cite{Iyer2019} use ad hoc functions, such as the minimum and average, to hybridize $y^*$ and $y^+$ to combine the strengths of various models. In contrast, the present approach does not rely on ad hoc blending functions and only uses $y^*$ since it has been established as the correct non-dimensionalization of various turbulent statistics \cite{Huang1995,Coleman1995}. 2) Volpiani et al. \cite{Volpiani2020} use data-fitting to determine the non-dimensionalizations of $y$ and $\pderi{U}{y}$. While this method performs well in the boundary layer cases which are used in their training database, we find that the model does not perform as well in channel and pipe flows.

A direct comparison reveals that the present transformation is more accurate than that obtained from the recent data-driven approach of Volpiani et al. \cite{Volpiani2020}, particularly in channel flows (as will be demonstrated next). Additional advantages of the present method are that it is based on physical arguments and uses the well-established semi-local non-dimensionalization of the wall-normal coordinate \cite{Huang1995,Coleman1995}. Unlike the present transformation and the Trettel-Larsson transformation, the data-driven approach does not recover the well-established viscous-sublayer transformation in the near-wall region \cite{Carvin1988}.

\begin{figure*}
\centering
  \includegraphics[width=0.32\linewidth]{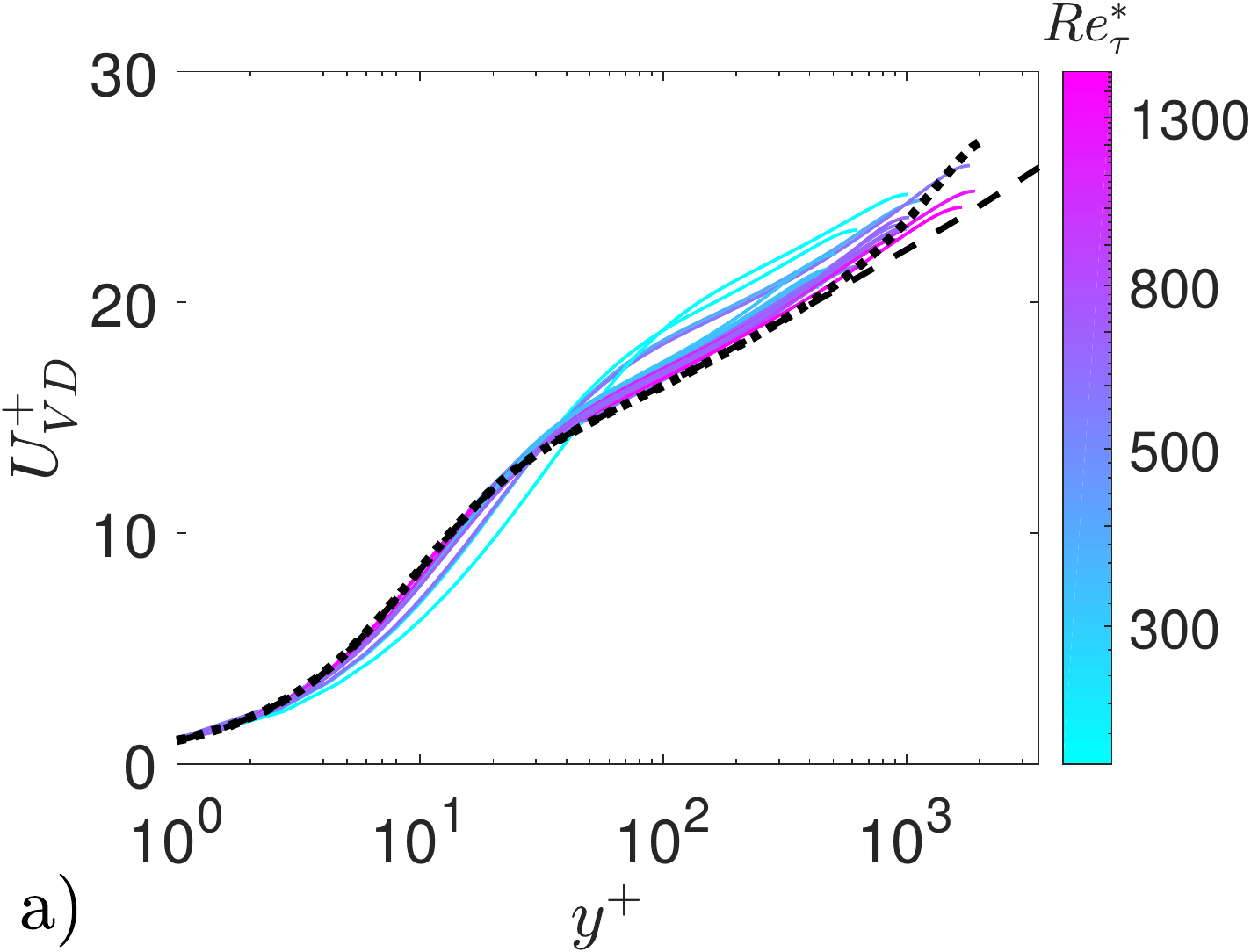}
  \includegraphics[width=0.32\linewidth]{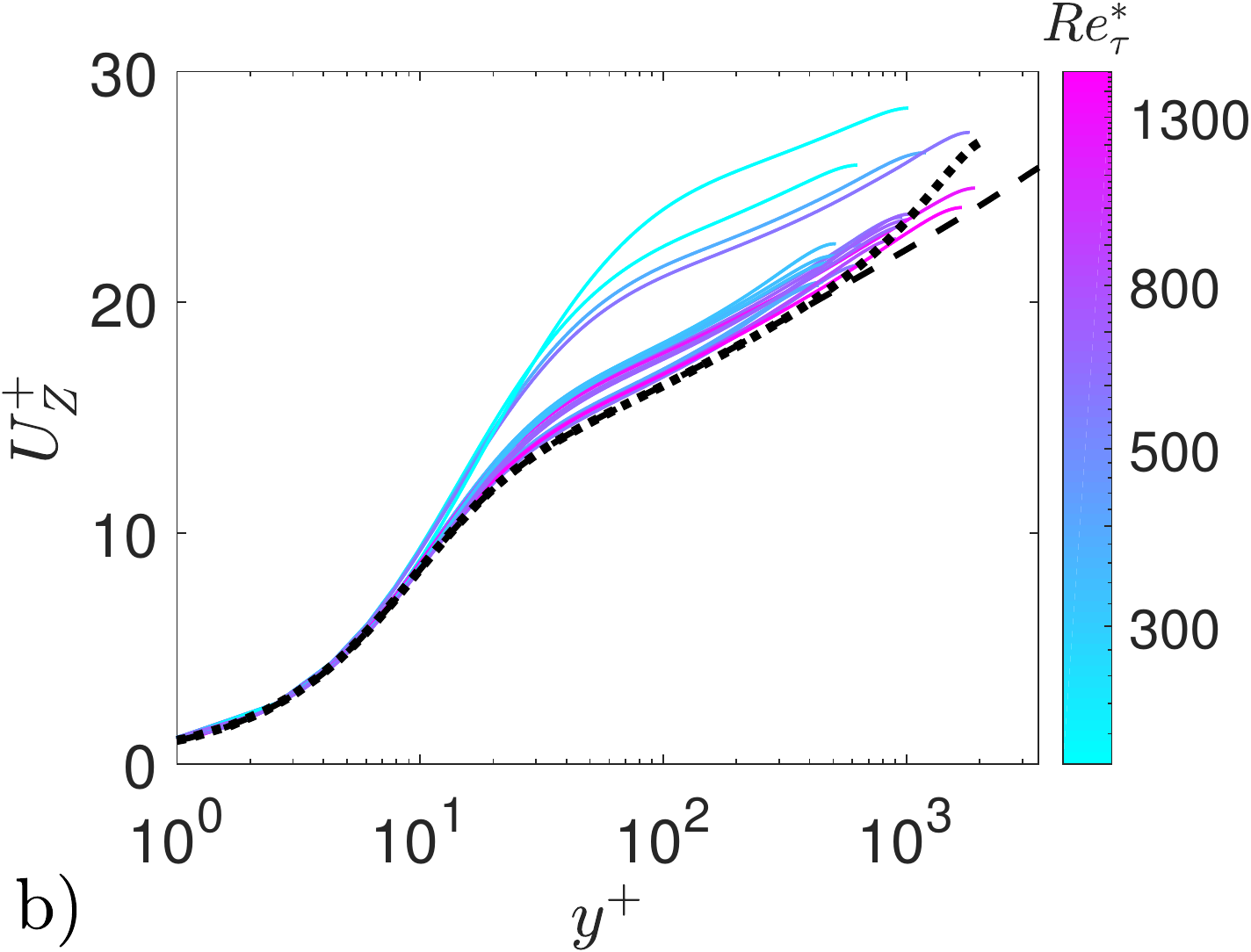}
  \includegraphics[width=0.32\linewidth]{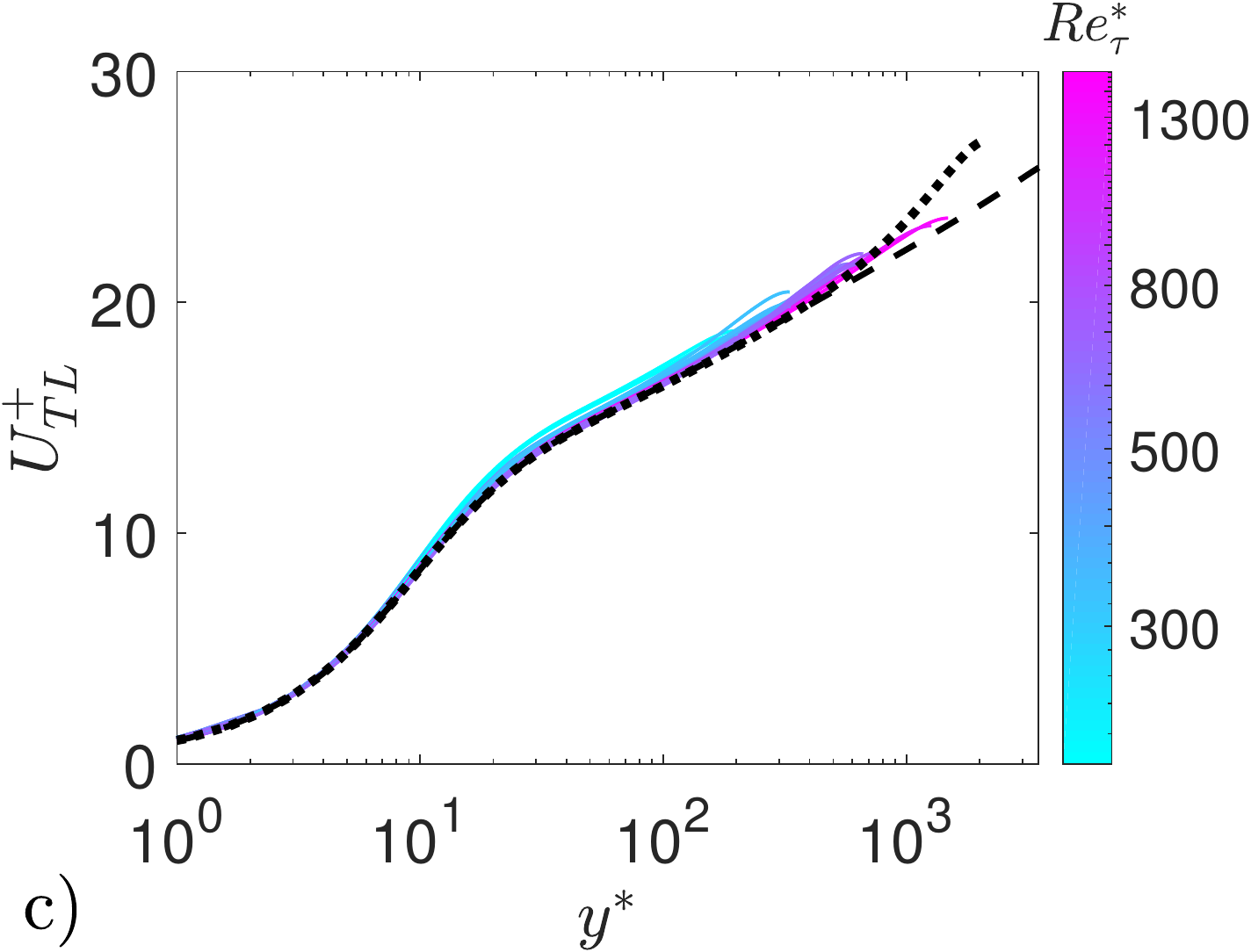}
  \includegraphics[width=0.32\linewidth]{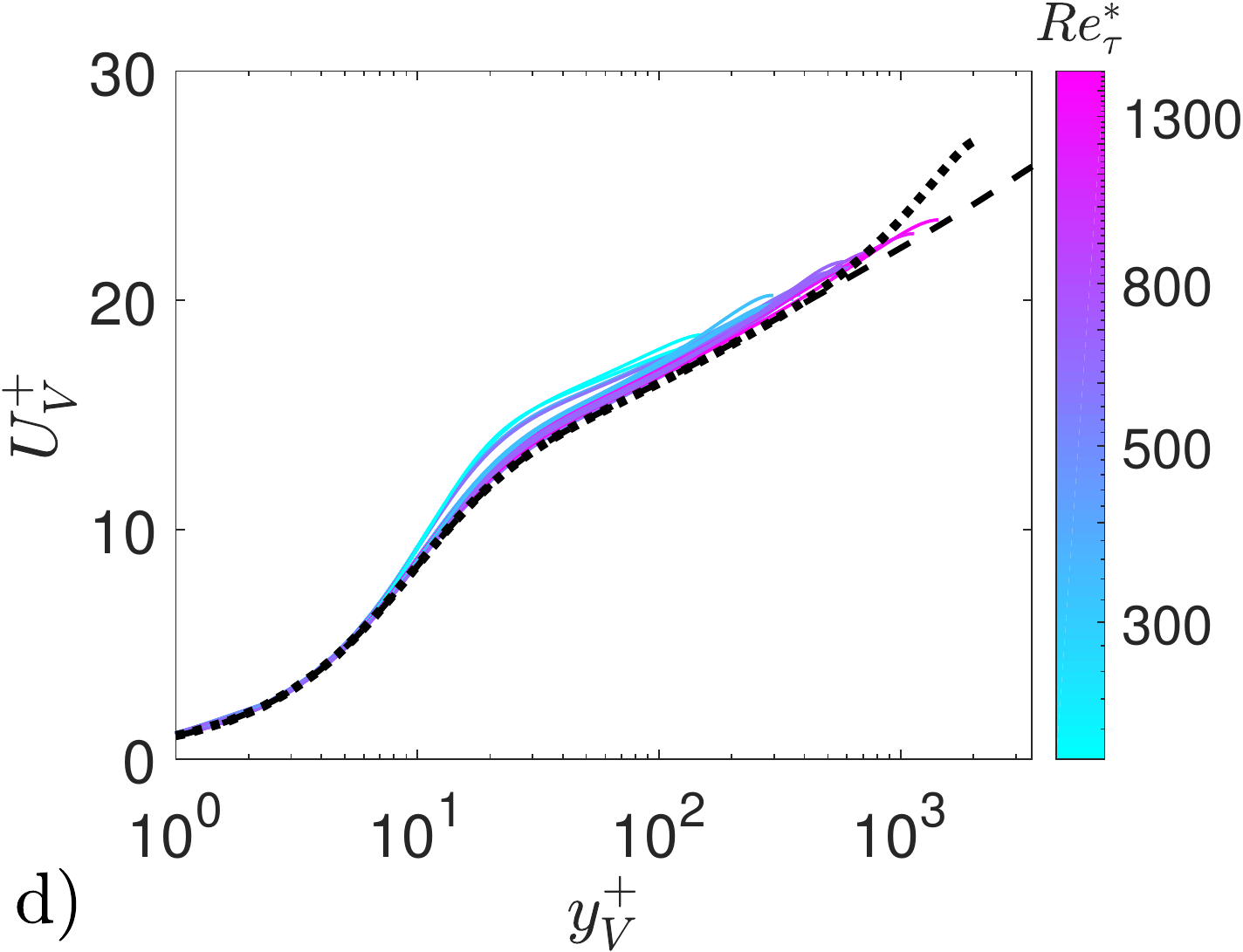}
  \includegraphics[width=0.32\linewidth]{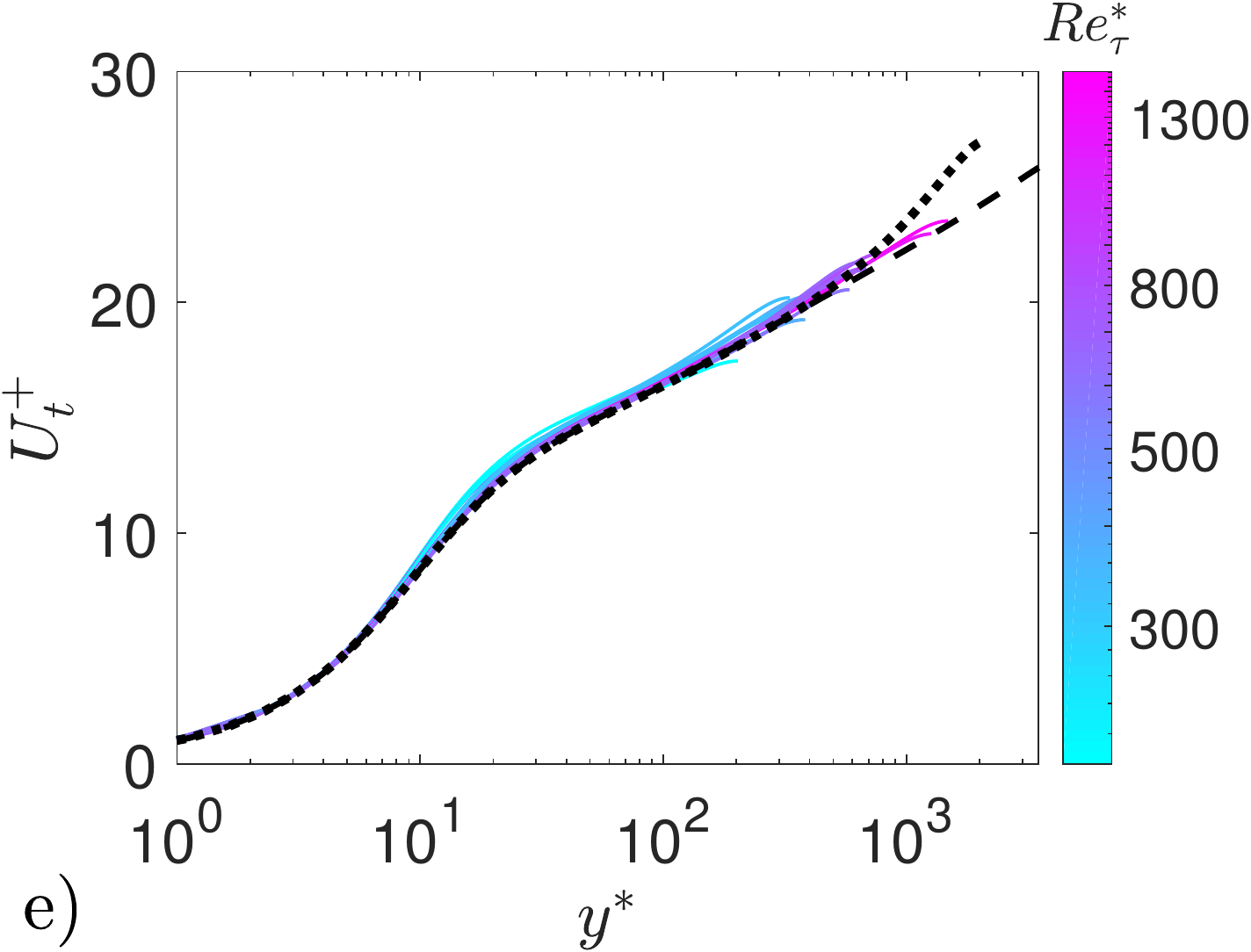}
\caption{Compressible velocity transformations versus the semi-local wall-normal coordinate for the van Driest \cite{VanDriest1951} (a), Zhang et al. \cite{Zhang2012} (b), Trettel-Larsson \cite{Trettel2016} (c), data-driven approach of Volpiani et al. \cite{Volpiani2020} (d), and the present total-stress-based  (e) transformations. The database includes: channel flows \citep{Modesti2016,Trettel2016,Yao2020} and pipe flows \citep{Modesti2019}. The line color indicates the semi-local Reynolds number $Re_\tau^* = \delta \sqrt{\tau_w \rho}/\mu$.  All data have $Re_\tau^*$ larger than 200. Incompressible channel data of Lee and Moser \cite{Lee2015} with $Re_\tau \approx 5200$ (black dashed lines) and the zero-pressure-gradient boundary layer data of Sillero et al. \cite{Sillero2013} with $Re_\tau \approx 2000$ (black dotted lines) are shown for reference.}
\label{fig:ChanPipe}
\end{figure*}

The transformed velocity profiles for the same compressible flows considered in Fig.~\ref{fig:all_cases} are presented here for three classes of flows separately. In Fig.~\ref{fig:ChanPipe}, five velocity transformations are applied to the channel and pipe flow data. The van Driest \cite{VanDriest1951} and Zhang et al. \cite{Zhang2012} transformations are the least accurate since they were developed for adiabatic boundary layers. At low Reynolds numbers, the Volpiani et al. \cite{Volpiani2020} transformation generates higher error than the Trettel-Larsson \cite{Trettel2016} and present transformations.

\begin{figure*}
\centering
  \includegraphics[width=0.32\linewidth]{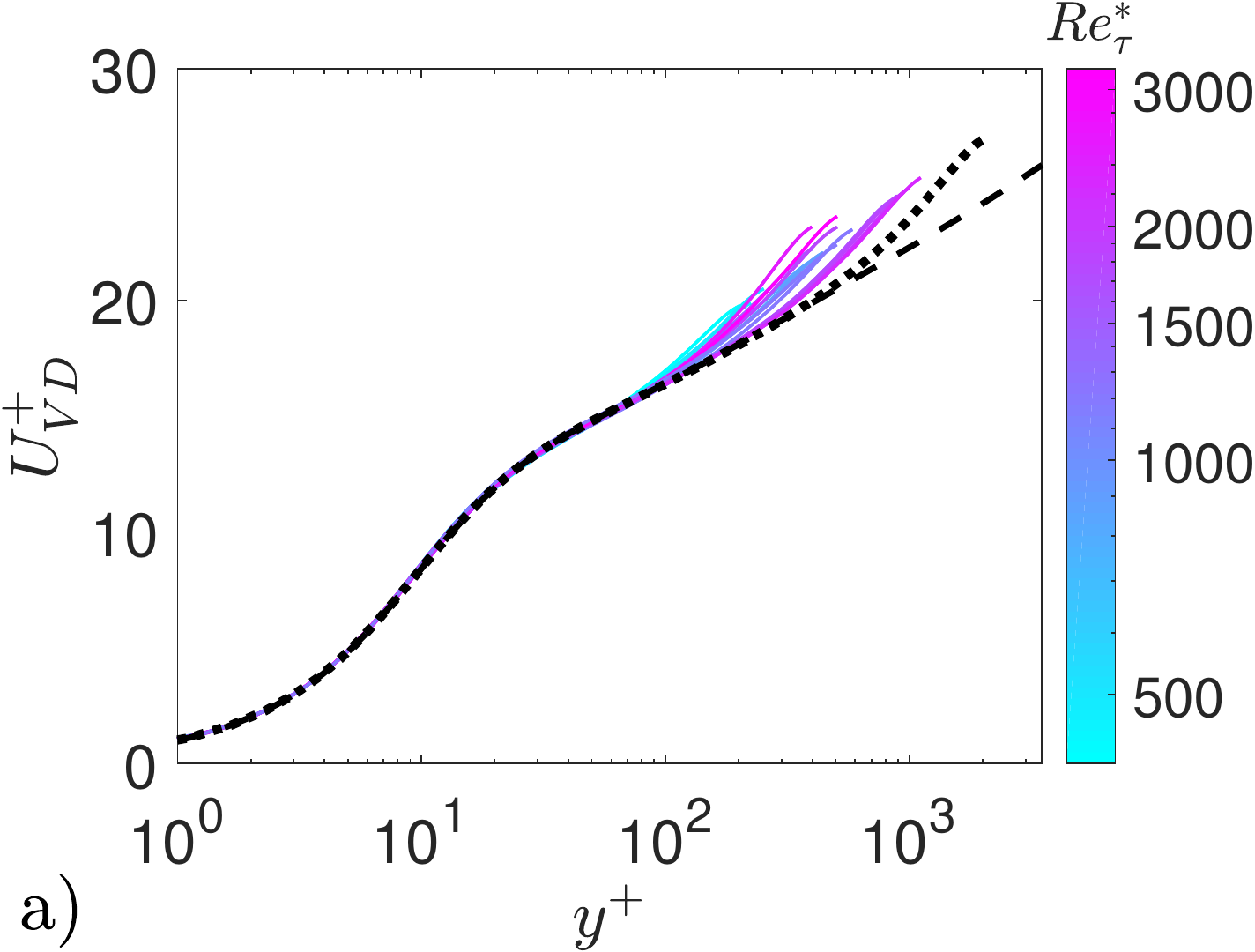}
  \includegraphics[width=0.32\linewidth]{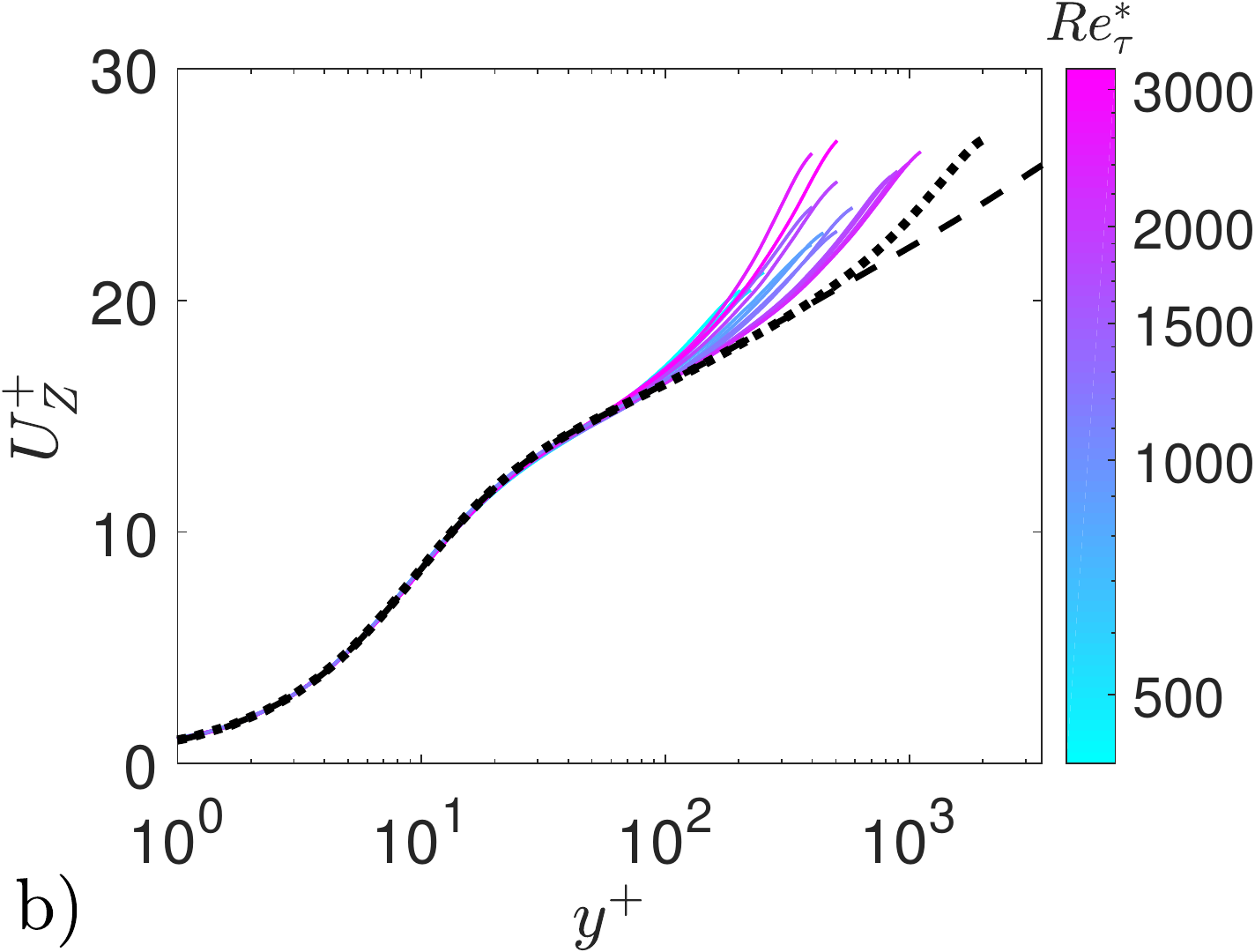}
  \includegraphics[width=0.32\linewidth]{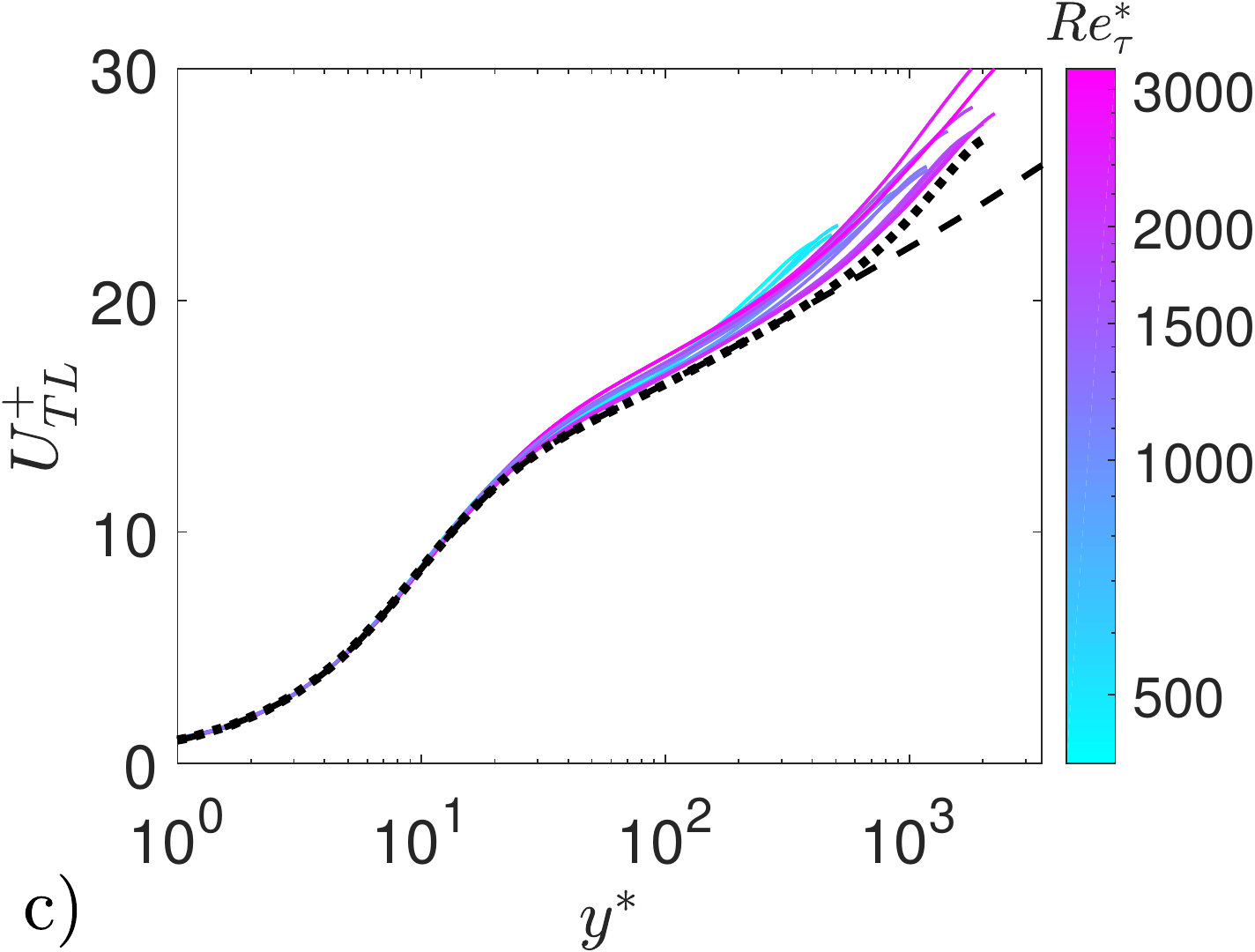}
  \includegraphics[width=0.32\linewidth]{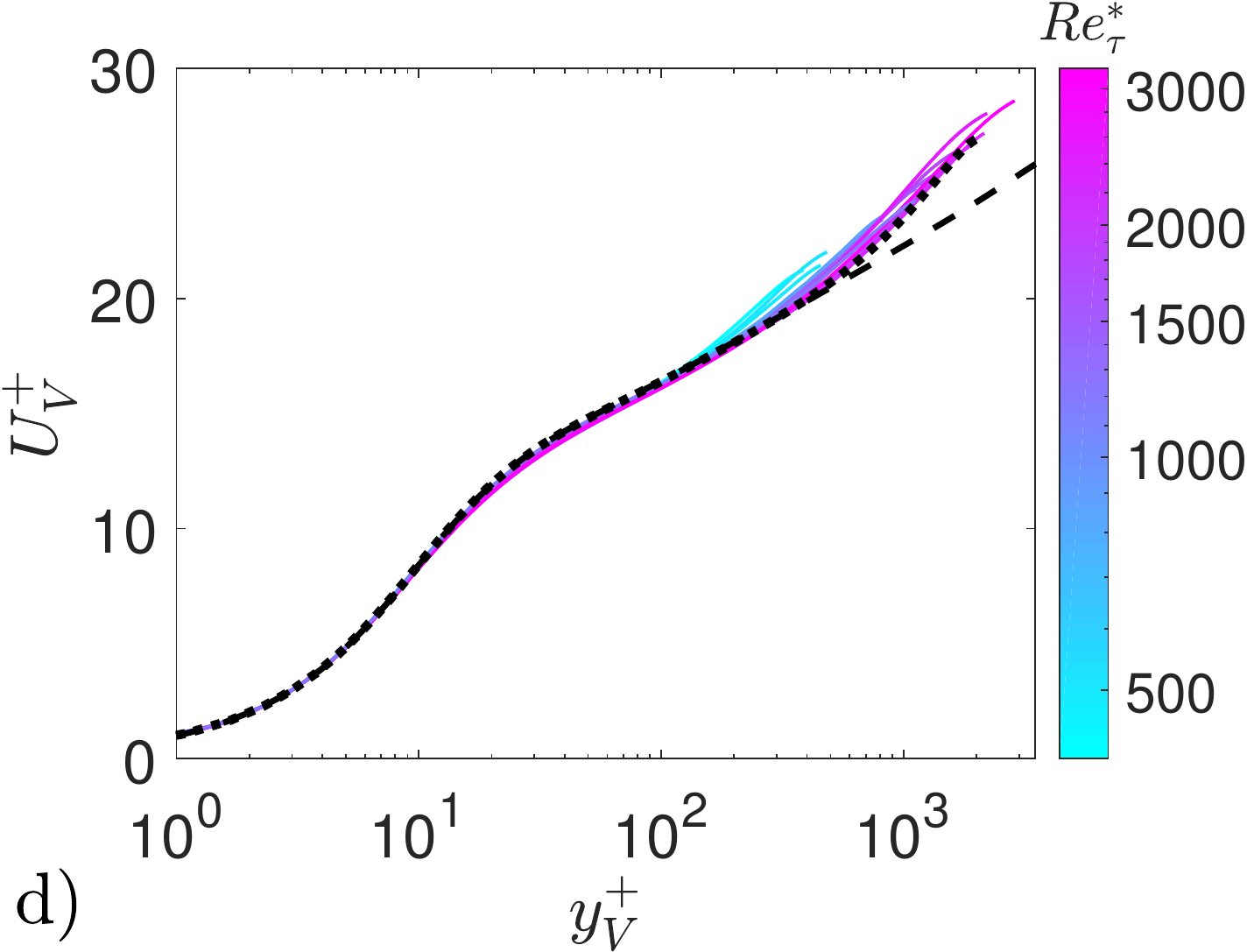}
  \includegraphics[width=0.32\linewidth]{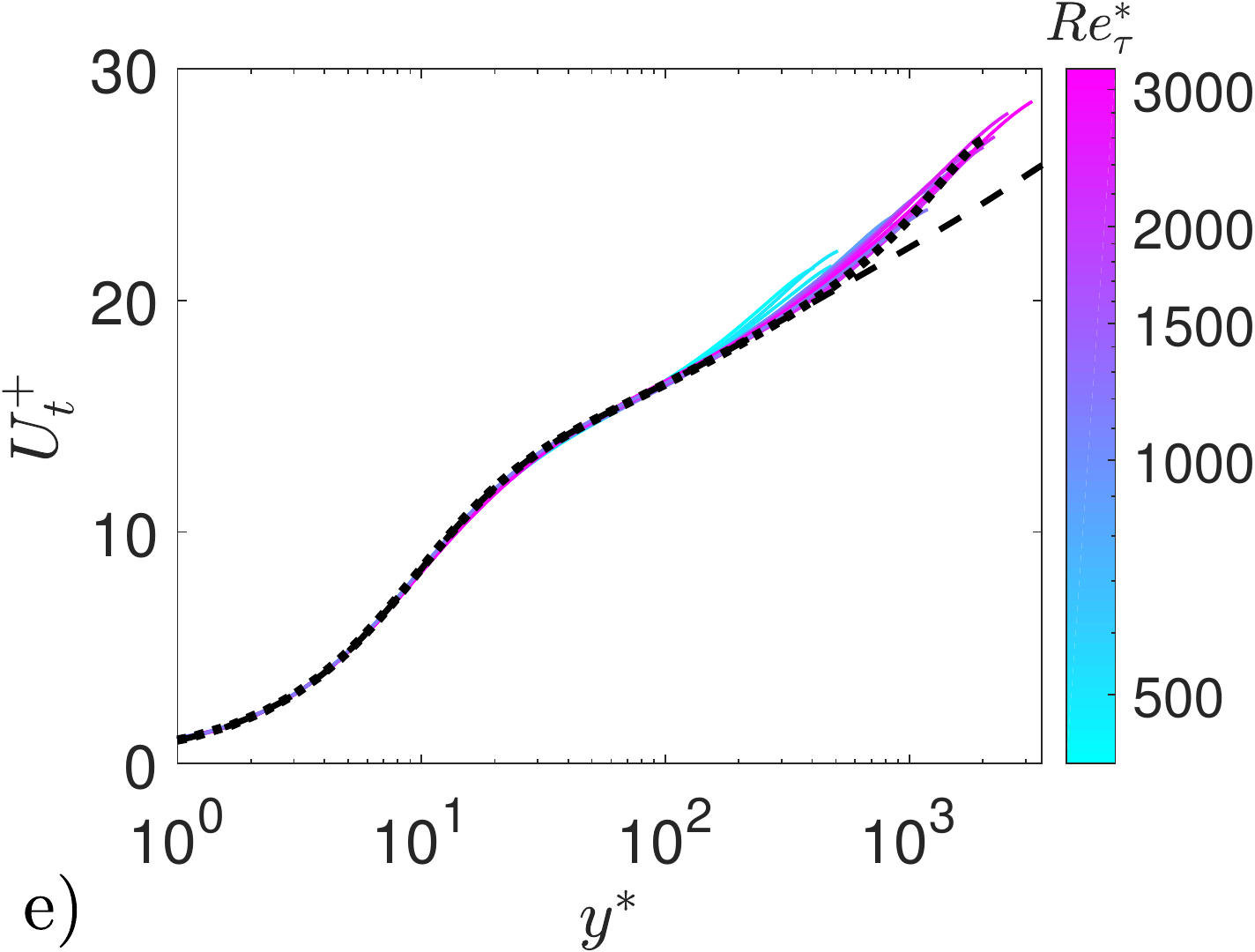}
\caption{Same as Fig.~\ref{fig:ChanPipe}, except that the transformations are applied to data from adiabatic boundary layers \citep{Pirozzoli2011,Zhang2018,Volpiani2018}.}
\label{fig:Adiabatic}
\end{figure*}

In Fig.~\ref{fig:Adiabatic}, the velocity transformations are applied to the adiabatic boundary layer data. The van Driest \cite{VanDriest1951} and Zhang et al. \cite{Zhang2012} transformed profiles depart from the incompressible reference data in the outer region, but this discrepancy in the wake region can be eliminated by comparing with the incompressible data at a matched $Re_\tau$, by similar arguments advanced by Modesti and Pirozzoli \cite{Modesti2016}. On the other hand, for the Trettel-Larsson \cite{Trettel2016} transformation, there is some scatter with regard to the log intercept and this scatter can not be attributed to the Reynolds number of the reference data. The Volpiani et al. \cite{Volpiani2020} and present transformations are accurate even at low Reynolds numbers.

\begin{figure*}
\centering
  \includegraphics[width=0.32\linewidth]{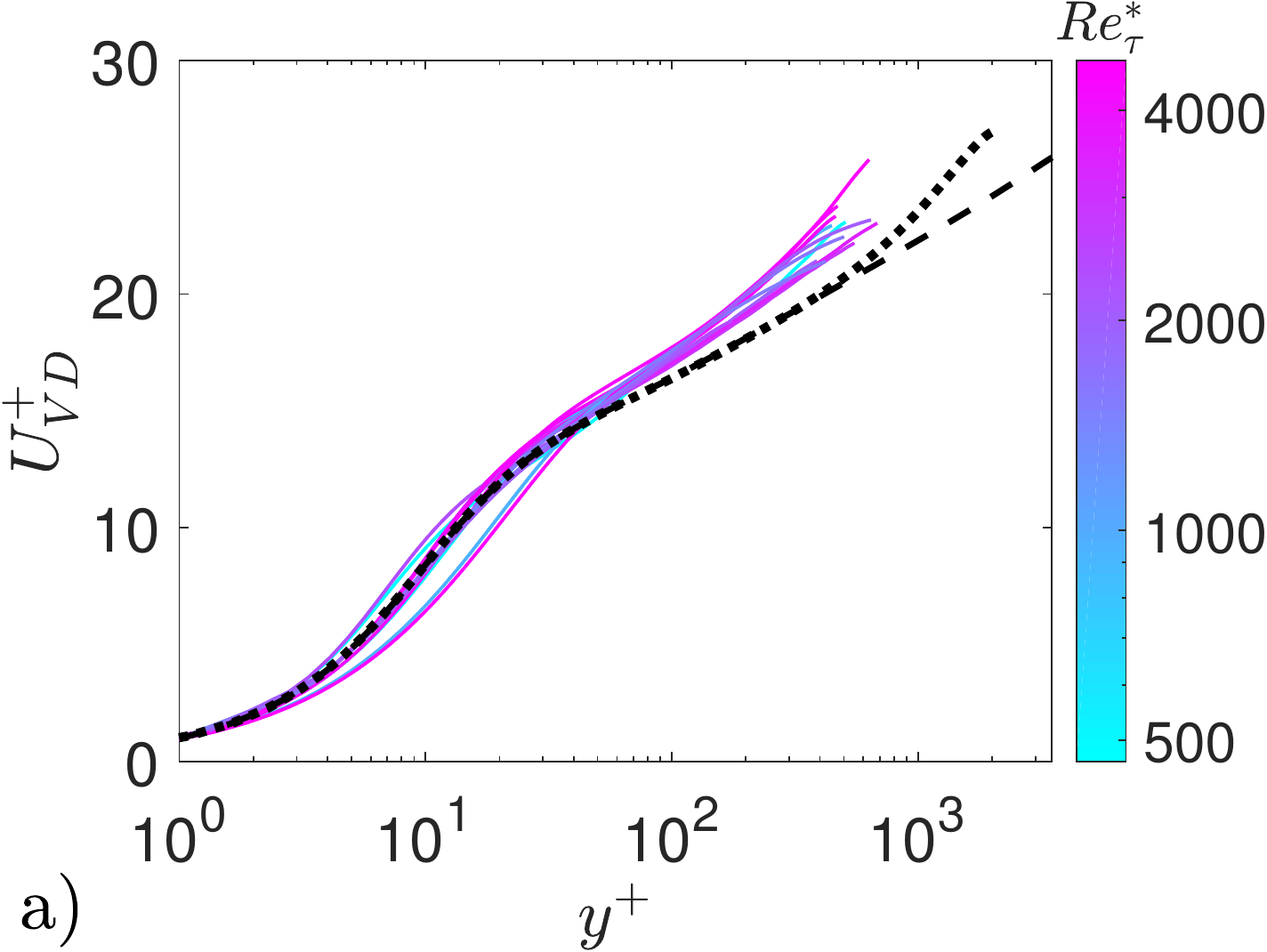}
  \includegraphics[width=0.32\linewidth]{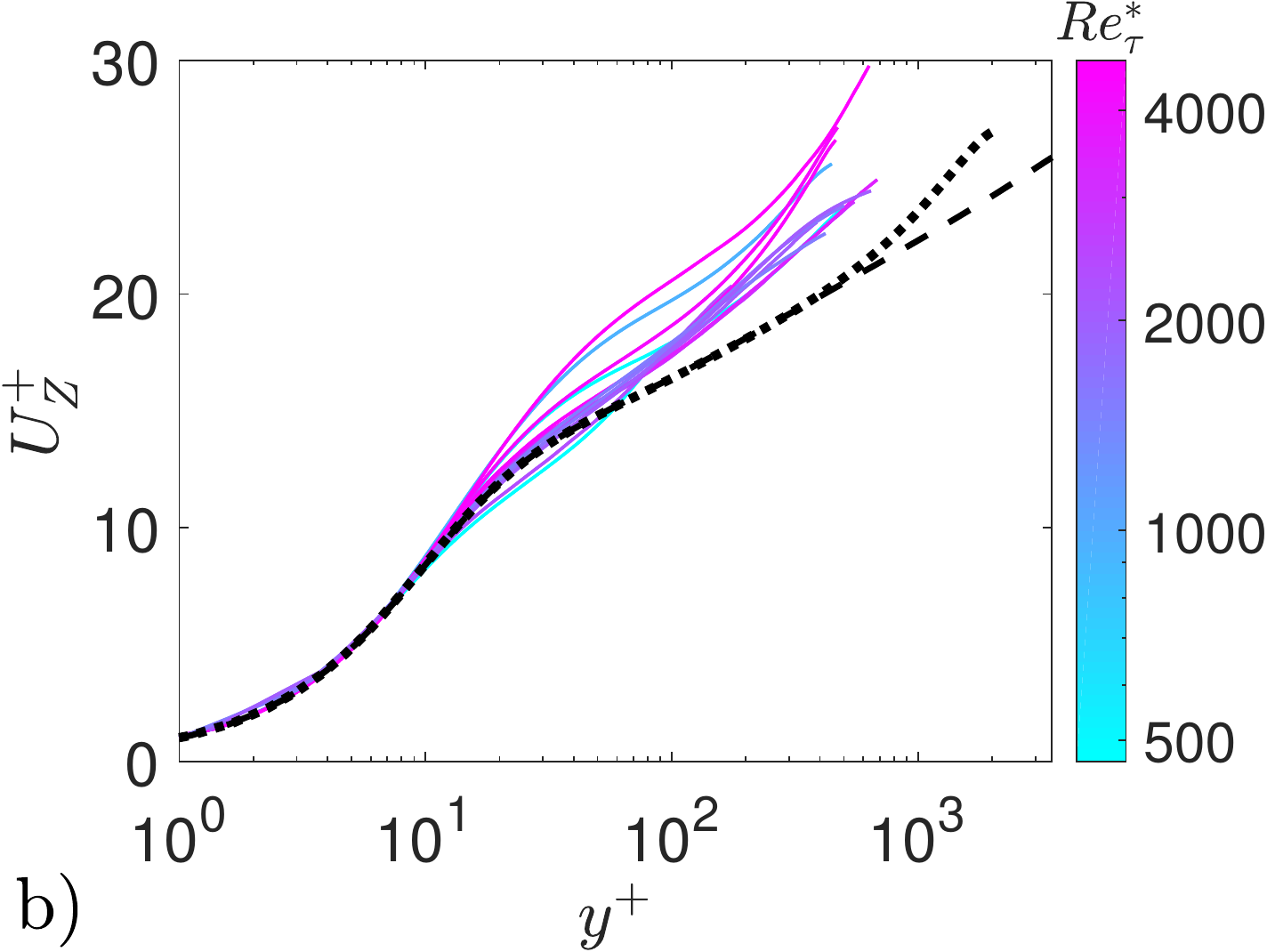}
  \includegraphics[width=0.32\linewidth]{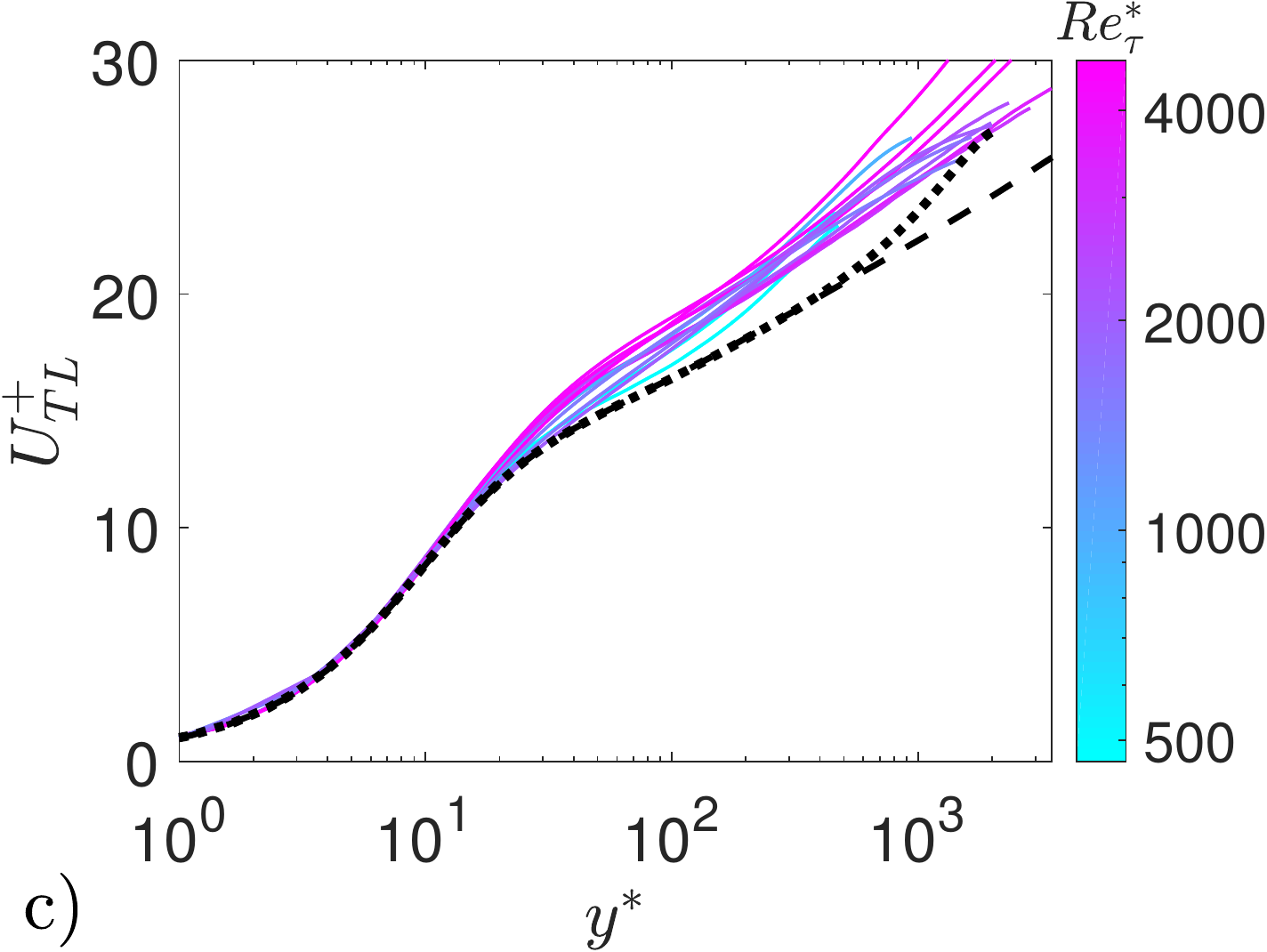}
  \includegraphics[width=0.32\linewidth]{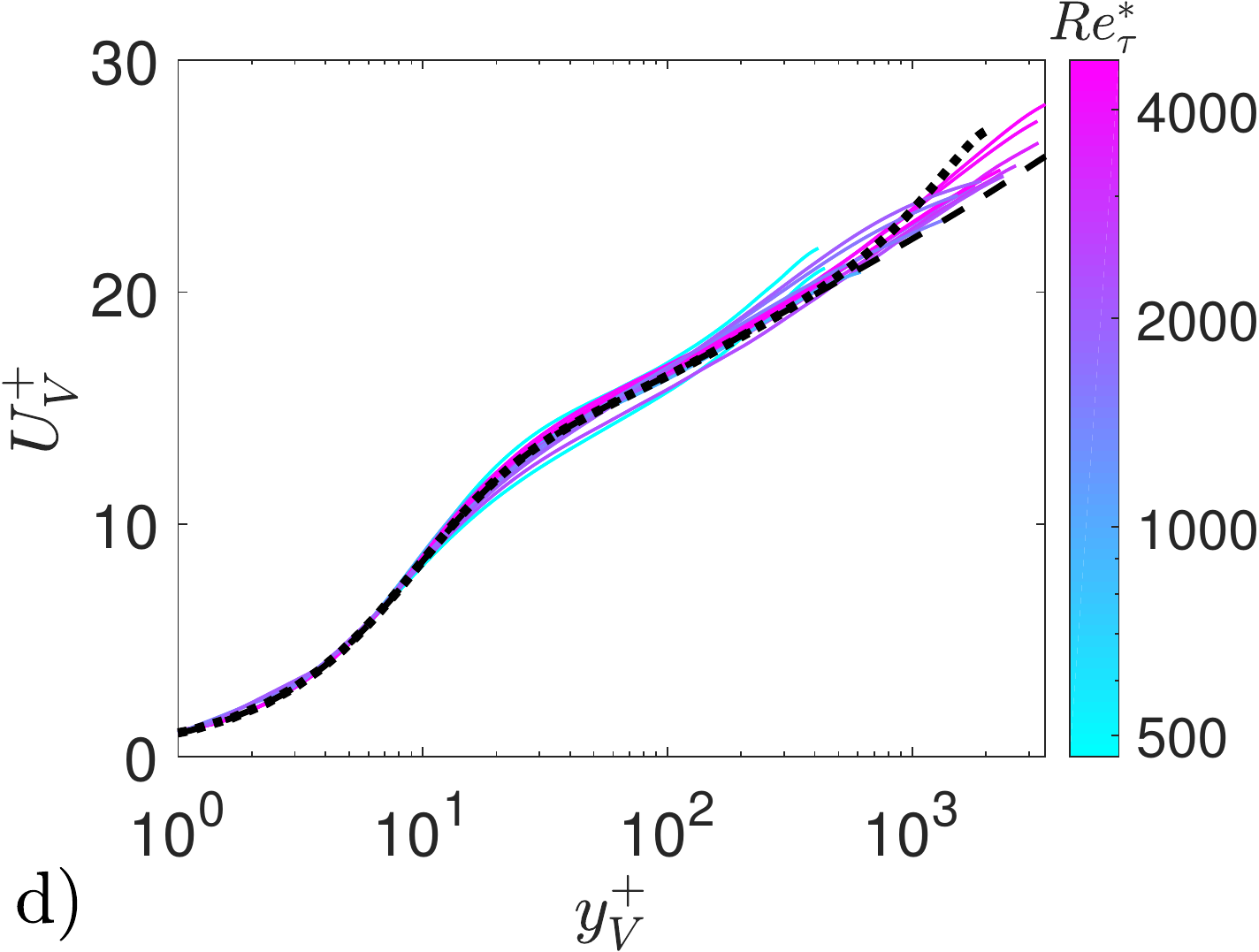}
  \includegraphics[width=0.32\linewidth]{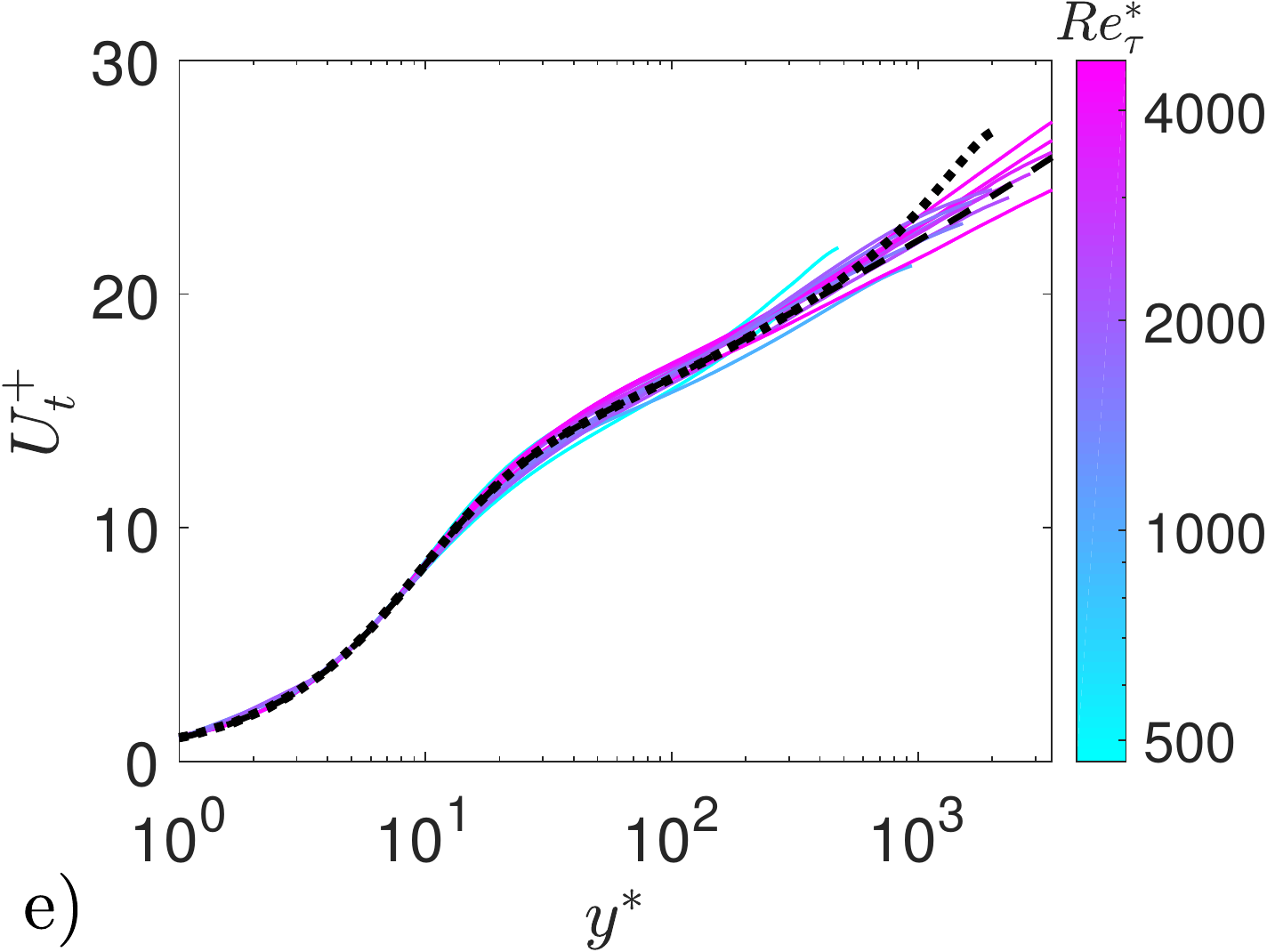}
\caption{Same as Fig.~\ref{fig:ChanPipe}, except that the transformations are applied to data from diabatic boundary layers (with both heated and cooled walls) \citep{Zhang2018,Volpiani2018,Volpiani2020a} and  a diabatic turbulent boundary layer downstream of the impingement of a shockwave on a laminar boundary layer \citep{Fu2021}.}
\label{fig:Diabatic}
\end{figure*}

In Fig.~\ref{fig:Diabatic}, the velocity transformations are applied to the diabatic boundary layer data. Similar to the results for channel and pipe flows, the van Driest \cite{VanDriest1951} and Zhang et al. \cite{Zhang2012} transformations are inaccurate since they were derived for adiabatic flows. The Trettel-Larsson \cite{Trettel2016} transformed profiles have log intercepts that do not agree with the incompressible reference data. This shift in the data is the result of the integrated overprediction of the mean shear documented in Fig.~\ref{fig:dU}(b). On the other hand, the transformations of Volpiani et al. \cite{Volpiani2020} and the present method are the most accurate for these flows.


Next, the three transformations with the lowest error are compared quantitatively.
The integrated percent error for the transformed velocity profiles can be defined as
\begin{equation}
    \epsilon = 100 \frac{\int_0^{100} |U_a^+ - U_I^+| dy_a}{\int_0^{100} U_I^+ dy_a},
\end{equation}
which is applicable for any given velocity transformation. Here, $U_a^+$ is the non-dimensional transformed compressible velocity profile, and $y_a$ is the non-dimensional wall-normal coordinate that corresponds to a specific transformation. $U_I^+$ denotes the incompressible reference velocity profile (non-dimensionalized with respect to wall units) for a channel flow at $Re_\tau = 5200$ \cite{Lee2015}. Note that for all transformations considered in this work $y_a \rightarrow y^+$ in the incompressible limit. The integration limits are fixed in transformation units as $y_a=[0,100]$ in order to include contributions from the viscous sublayer, the buffer layer, and the log region, and to avoid diluting any near-wall errors in the high-Reynolds-number cases.

\begin{figure*}
\centering
  \includegraphics[width=1\linewidth]{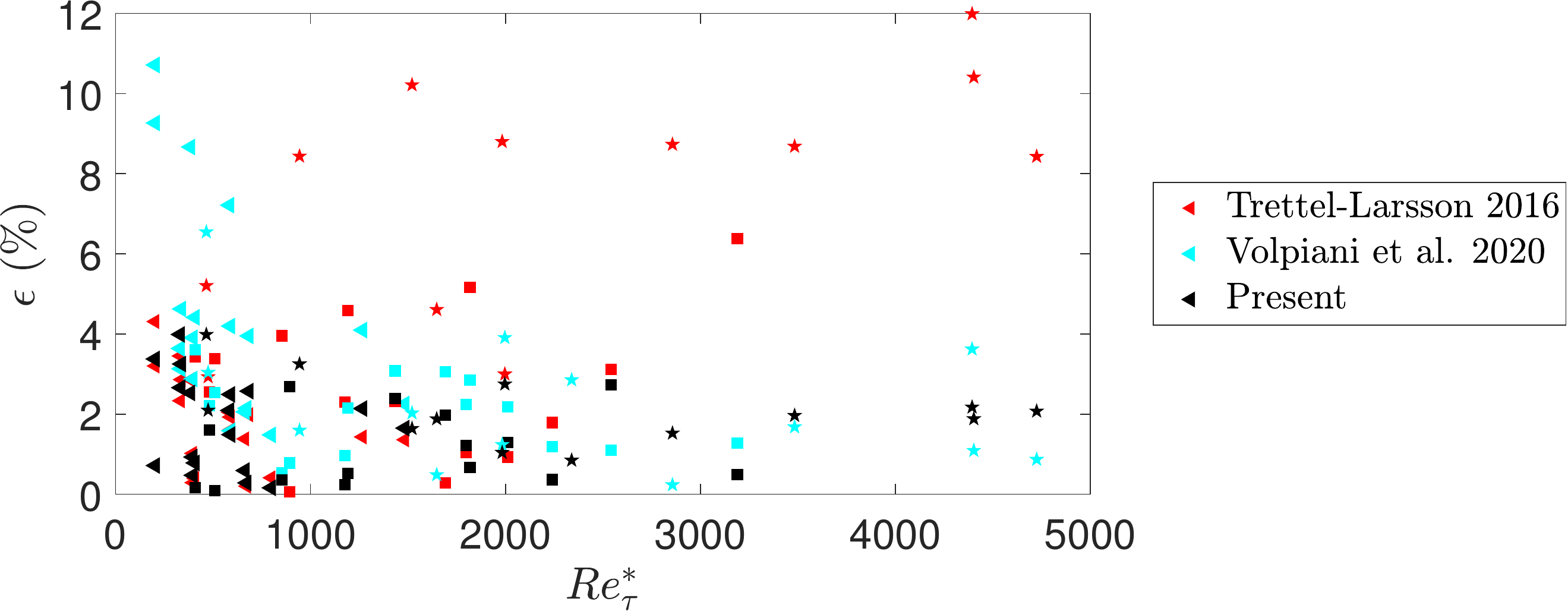}
\caption{Integrated percent error in the transformed velocity profiles according to the Trettel-Larsson \cite{Trettel2016} (red), Volpiani et al. \cite{Volpiani2020} (cyan), and present (black) transformations computed with respect to the incompressible reference \cite{Lee2015}.  All data have $Re_\tau^*$ larger than 200. Symbol shape indicates the flow type, i.e., triangles denote channel and pipe flows, squares adiabatic boundary layers, and pentagrams diabatic boundary layers.}
\label{fig:err}
\end{figure*}

In Fig.~\ref{fig:err}, the integrated percent errors for the transformed velocity profiles are shown for the three most accurate transformations: the Trettel-Larsson \cite{Trettel2016}, Volpiani et al. \cite{Volpiani2020} (data-driven approach), and present transformations. The invoked compressible data is the same as in Fig.~\ref{fig:all_cases}, i.e., adiabatic boundary layers \citep{Pirozzoli2011,Zhang2018,Volpiani2018}, diabatic boundary layers (with both heated and cooled walls) \citep{Zhang2018,Volpiani2018,Volpiani2020a}, channel flows \citep{Modesti2016,Trettel2016,Yao2020}, pipe flows \citep{Modesti2019}, and a diabatic turbulent boundary layer downstream of the impingement of a shockwave on a laminar boundary layer \citep{Fu2021}.

The Trettel-Larsson \cite{Trettel2016} transformation generates the largest errors in diabatic boundary layers, and what is worse is that the errors do not diminish at high semi-local Reynolds numbers. The transformation of Volpiani et al. \cite{Volpiani2020} generates the largest errors in channels and pipes (which generally have low semi-local Reynolds numbers due to the center-line temperature peak). On the other hand, the present approach has the lowest percent error for most of the cases with errors typically less than three percent.

As shown in Fig.~\ref{fig:err}, the performance of the present transformation does not appear to degrade at high Reynolds numbers. However, until additional simulation data becomes available, some caution is merited in the deployment of this transformation for cases that have not been tested, such as in higher Reynolds number flows or flows with strong pressure gradients.

Forward velocity transformations attempt to collapse compressible velocity profiles to a universal incompressible velocity profile. On the other hand, an inverse transformation can, in principle, be used to map a tabulated incompressible profile to a compressible flow with arbitrary freestream Mach number. The simplest of the considered transformations in this work, that of van Driest \cite{VanDriest1951}, has an analytical inverse form if the Prandtl number is assumed to be unity, the constant-stress-layer assumption is invoked, and the gas is assumed to be perfect \cite{White2006}. However, for the present and other recent transformations, analytical inverse transformations are not apparent and are not pursued further in this work.

\section*{Conclusions}

In summary, the appropriate mapping between the compressible and incompressible mean velocity profiles of wall-bounded turbulent flows must account for distinct effects of compressiblity on the viscous stress and turbulent shear stress. The present transformation is fundamentally different from prior approaches, which have assumed that compressibility affects the viscous stress and Reynolds shear stress through only one mechanism. Specifically, the present approach treats the viscous stress by accounting for mean property variations with the semi-local non-dimensionalization \citep{Huang1995,Coleman1995,Trettel2016,Patel2016} and treats the Reynolds shear stress to maintain the approximate equilibrium of turbulence production and dissipation.

Compared to existing transformations \cite{Zhang2012,Trettel2016}, the proposed method is unique in its general applicability to different types of wall-bounded turbulence. Specifically, based on the proposed transformation, we have shown that compressible velocity profiles successfully collapse to high Reynolds number incompressible data for a wide range of flows including heated, cooled, and adiabatic boundary layers; fully developed channel and pipe flows; and turbulent boundary layers downstream of the impinging shock waves.

\acknow{K.P.G. acknowledges support from the National Defense Science and Engineering Graduate Fellowship and the Stanford Graduate Fellowship. L.F. and P.M. are supported by NASA grant NNX15AU93A and by funding from Boeing Research \& Technology. We are grateful to Dr.~Andrew Trettel for assistance in procuring data. We wish to acknowledge helpful feedback from Dr. Sanjeeb T. Bose, Dr. Javier Urzay, and Dr. W. H. Ronald Chan.}

\showacknow{} 

\bibliography{comp_database.bib,comp_vel_transf.bib,comp_wall.bib,incomp_database.bib,wall_model.bib,1AAA_mine.bib}
	
\end{document}